\documentclass[useAMS,usenatbib]{mn2e}
\usepackage{graphicx,bm,amssymb,amsmath}
\newcommand{\lcdm}{$\Lambda$CDM}
\newcommand{\conc}{c_{\Delta}}
\newcommand{\Omk}{\Omega_{\mathrm{k}}}
\newcommand{\OmL}{\Omega_{\mathrm{\Lambda}}}
\newcommand{\OmM}{\Omega_{\mathrm{M}}}
\newcommand{\OmMo}{\Omega_{\mathrm{M},0}}

\newcommand{\OmLo}{\Omega_{\mathrm{\Lambda},0}}
\newcommand{\msol}{\mathrm{M}_{\odot}}

\title[Raytracing in 3D]{Gravitational Lensing with Three-Dimensional Ray Tracing}

\author[M. Killedar et al.]{
M.~Killedar$^{1,2}$\thanks{
Research undertaken as part of the Commonwealth Cosmology Initiative (CCI: www.thecci.org), an international collaboration supported by the Australian Research Council (ARC).
E-mail:
killedar@oats.inaf.it (MK); 
lasky@tat.physik.uni-tuebingen.de (PL); 
}, 
P.~D.~Lasky$^{3,4}$\footnotemark[1], 
G.~F.~Lewis$^{1}$ \& C.~J.~Fluke$^{4}$  \\\
$^{1}$Sydney Institute for Astronomy, School of Physics, The University of Sydney, NSW, 2006, Australia\\
$^{2}$Dipartimento di Fisica dell'Universit\`{a} di Trieste, Sezione di Astronomia, Via Tiepolo 11, I-34131 Trieste, Italy\\
$^{3}$Theoretical Astrophysics, Eberhard Karls University of T\"{u}bingen, T\"{u}bingen 72076, Germany\\
$^{4}$Centre for Astrophysics and Supercomputing, Swinburne University of Technology, PO Box 218, Hawthorn, 3122, Victoria, Australia}

\begin{document}
\bibliographystyle{scemnras} 

\date{Accepted 2011 October 15. Received 2011 October 04}
%\date{}                                           % Activate to display a given date or no date

\pagerange{\pageref{firstpage}--\pageref{lastpage}} \pubyear{2011}

\maketitle

\label{firstpage}

%--------------------------------------------------------------------------------------------------------------------------------- 

\begin{abstract} 
High redshift sources suffer from magnification or demagnification due to weak gravitational lensing by large scale structure. One consequence of this is that the distance-redshift relation, in wide use for cosmological tests, suffers lensing-induced scatter which can be quantified by the magnification probability distribution. Predicting this distribution generally requires a method for ray-tracing through cosmological N-body simulations. However, standard methods tend to apply the multiple thin-lens approximation. In an effort to quantify the accuracy of these methods, we develop an innovative code that performs ray-tracing without the use of this approximation. The efficiency and accuracy of this computationally challenging approach can be improved by careful choices of numerical parameters; therefore, the results are analysed for the behaviour of the ray-tracing code in the vicinity of Schwarzschild and Navarro-Frenk-White lenses. Preliminary comparisons are drawn with the multiple lens-plane ray-bundle method in the context of cosmological mass distributions for a source redshift of $z_{s}=0.5$. 
\end{abstract}


%---------------------------------------------------------------------------------------------------------------------------------
\section{Introduction}

Weak gravitational lensing of distant galaxies by large-scale structure leads to the distortion of their images. In an inhomogeneous universe, each line of sight probes a slightly different integrated mass and is sheared differently; overdense regions result in magnification and underdense regions cause demagnification. The result is that for a given redshift $z$, a source is magnified (or demagnified) by some amount $\mu$, {\it relative to the case of a perfectly homogeneous universe on all scales}. This magnification has an associated probability distribution that reflects the existence of structure on a range of scales.

Large surveys for galaxies, quasars and supernovae suffer a two-fold {\it magnification bias} as a result of the same phenomenon. When a source is magnified, its total surface area on the sky is increased; consequently, for a given area of sky observed, the region of the source plane being sampled is decreased. On the other hand, magnification will push otherwise too-faint sources above the observational threshold in flux-limited surveys \citep{C81, TOG84}; this is particularly important for optically-selected quasar surveys \citep{P82, BL96, W11}.
Together, these two effects have counteracting influences on source counts, but generally do not cancel each other out. Additionally, for sources that are demagnified the effect is reversed; indeed, the majority of sources are de-magnified relative to the case of a perfectly homogeneous universe. 
The net effect on source counts, observed luminosity functions, source redshift distributions and any resulting bias depends on selection procedures, intrinsic source properties and the probabilistic lensing effect \citep{DR74, P82, LB00}. 
The intrinsic source luminosity function and magnification probability distribution function (hereafter $\mu$PDF) are unknowns; the former is generally modelled as a Schechter function \citep{S76}. An appropriate model for the latter --- which may be produced by ray-tracing through N-body simulations --- is the subject of the present study (see Sec.~\ref{mpdf}). \citet{HMF00} and \citet{JL11} have demonstrated that a power-law tail in the $\mu$PDF significantly changes the shape of the bright end of the luminosity function and generates a considerable magnification bias for high-redshift sources.

Gravitational lensing causes magnification or demagnification of Type Ia supernovae (SNIa), resulting in a scatter in the inferred distance-redshift distribution, particularly for high-redshift sources \citep{J08}. 
The lensing-induced scatter introduces a bias and uncertainty in the inferred values of the matter density parameter, $\OmMo$, the deceleration parameter, $q_{0}$ and the dark energy equation of state, $w$ \citep{W97, A06, J08}. For example, in the case of the $z=1.7$ supernova SN1997ff, the estimated magnification would easily bridge the gap between evidence supporting a \lcdm~universe and that supporting an empty universe. The lensing effect can be `averaged out',  for supernovae samples, by combining measurements from a sufficiently large number of sources at similar redshifts  \citep{HW98, HL05}.
However, a similar scatter will be induced by lensing in the Hubble diagram constructed from measurements of gravitational waves, the so-called standard sirens \citep{M93,HH05, JGM07}.  Efforts are now underway to provide an appropriate method for correcting the observed brightnesses of individual objects for magnification \citep[e.g.][]{G06,JMS09,S10,HGK11}. By accounting for the non-Gaussian shape of the $\mu$PDF, one is best able to correct for the effects of lensing statistically \citep{HHC10, SH11}.

There are many approaches to determining the $\mu$PDF for a given cosmological model. 
The optical scalar equations \citep{S61}, which describe the evolution of the cross-section of a small beam of light, have been applied to various mass distributions leading to useful redshift-distance relations in the limiting case when the line of sight is far away for inhomogeneities \citep{K69,DR74,D77}. 
The equations have also been applied to an infinitesimal beam transported through a generalized inhomogeneous universe that, on average, satisfies the Friedmann-Lema\^{i}tre-Robertson-Walker (FLRW) geometry in order to re-derive the maximal angular diameter distance \citep{SSE94}.
A simple integration of the optical scalar equations would be possible in the case of a known metric, but that is the crux of the problem - the local metric for an inhomogeneous universe has no general (analytic) solution. Crude model universes incorporate inhomogeneities and describe their effects on light propagation \citep[e.g.][]{DR72,L88,FS89}.
However, when the cosmological structure being probed is highly non-linear and lines of sight have small impact parameters, studying light deflection requires the use of numerical techniques.

The non-linear and hierarchical growth of the large-scale structure is generally modelled by cosmological N-body simulations, with the propagation of light and its subsequent lensing computed by ray-tracing through these simulations. Some methods consider only one dominant lens in each line of sight; the lensing object is thin compared to the distances between observer, lens and source - this is known as the thin-lens approximation. However, multiple lenses coincidental along the line of sight may be responsible for a lensing event \citep{WBO05, DO06}. The existence of multiple lens plane also accounts for the findings of \citet{P01}: that the magnification probability, $P(\mu>1)$ is mostly independent of the parameter that quantifies the normalisation of the matter power spectrum, $\sigma_{8}$.
\citet{R92} identified a feature of the $\mu$PDF: a `bump' that was only evident when considering two-dimensional projections of matter; they speculated that changes to the caustic structure resulting from the projection was somewhat responsible. The presence of this caustic-induced bump was confirmed by \citet{L97}, who derived a semi-analytic expression for this feature in the limit of low-optical depth.
Therefore, in the context of cosmological structure, the thin-lens approximation should be replaced, at the very least, by the multiple-lens plane approach (see Sec.~\ref{multilensplane}), where large volumes of matter are projected onto a series of lens planes \citep{BN86, K87} . 
The ray-shooting method, developed by \citet{KRS86,SW88,W90,WCO98}, embodies this approach. 
Subsequent introduction of tree-methods to measure the deflection angle at each lens-plane have produced efficient algorithms. 
Early ray-shooting techniques were applied to a random distribution of point masses, which therefore did not incorporate the intricacies of clustering properties within various cosmological models \citep{SW88,PW89,LP90}. Later studies coupled the multiple lens-plane approach with mass distributions taken from N-body simulations to study the effect and magnitude of gravitational lensing \citep[e.g.][]{J90}. 

While it is tempting to continue using the previous methods because of their simplicity, one must first quantify the effect that the various approximations have on the resulting predictions. Galaxy redshift surveys such as the CfA survey \citep{D82} and the 2dFGRS \citep{Cole05} have revealed that elongated structures exist in the form of filaments, which stretch across large voids between galaxy clusters. These are also evident in cosmological simulations. If such filaments are projected onto lens planes, the resulting magnification would be overestimated.
Though earlier computational limitations, such as memory and processing, warranted the need for these simplifications, we are now entering an era where more accurate methods are within our reach.

Only a few studies have numerically integrated the null geodesic equations from observer to source. The earliest of these studies assumed metrics that were approximated via a simplified model for inhomogeneities and derived distance-redshift relations that were compared to the Dyer-Roeder approximations \citep{FS89,WT90}. \citet{KFT90}, drew attention to the spread in angular diameter distances given a small enough beam-size and the fact that the average value was consistently lower than the solution for a homogeneous universe.
\citet{T98} numerically integrated the null geodesic equations through N-body simulations, albeit at low-resolution, finding that various cosmologies exhibited differences in angular diameter distances.
\citet{CBT99} developed a three-dimensional algorithm for modelling weak gravitational lensing; comparing the two- and three-dimensional shear, they found that the projection of structure on 100 h$^{-1}$Mpc scales led to errors of up to 9 per cent depending on the redshift of the lens box. \citet{VW03} performed three-dimensional ray-tracing by computing the deflection angle for rays many times along their path, advancing them in step-sizes of $L_{\mathrm{Box}}/N_{g}$ along the line of sight. In their study, they used a relatively small number of grid points along the line-of-sight $N_{g}=32$, such that the step-size was approximately 10 h$^{-1}$Mpc, but boxes of half that size were used for numerical tests. They found that their convergence maps were unchanged to 0.1 per cent. \citet{WH00} run simulations of structure formation while simultaneously propagating photons through said simulation; they were able to produce convergence fields from the light rays they follow. 

What has been missing thus far is a one-to-one comparison of results that take a multiple-lens approach and a direct numerical integration of the null geodesic equations; this comparison is carried out in the present work.
In Section~\ref{gravlensing}, the basic theory and notation of gravitational lensing and the relevant statistics are presented.
We introduce the method of numerically integrating the null geodesic equations in Section~\ref{method} with specific reference to its multiple lens plane counterpart; in Section~\ref{models}, we establish the accuracy and limitations of the method. 
We describe the simulations used for modelling the cosmological mass distribution and compare statistical predictions determined by the two methods in Section~\ref{results}. 
Finally, a summary of our findings is presented in Section~\ref{summary}.

%---------------------------------------------------------------------------------------------------------------------------------
\section{Gravitational Lensing}\label{gravlensing}
Here, we introduce the conventional notation for the components that describe the effect of gravitational lensing on the shape of a beam to first order. Each background source experiences lensing in the form of convergence, $\kappa$, and shear, $\boldsymbol{\gamma}$. Convergence is the isotropic Ricci focusing of a beam due to enclosed matter, while shear is the tidal stretching of the beam along a particular axis due to asymmetric matter distribution; together, they serve to increase the area of the source on the sky, resulting in its magnification, due to the conservation of surface brightness. 
The two-dimensional effective lensing potential, $\psi_{2D}$, is given by
\begin{equation}\label{potential}
	\psi_{2D}(\mathbf{x}) = \frac{1}{\pi}\int_{\mathbb{R}^{2}} d^2x'\kappa(\mathbf{x}')\mathrm{ln}\left|\mathbf{x}-\mathbf{x'}\right| .
\end{equation}
Here $\mathbf{x}$ is a dimensionless vector formed by scaling the image position, $\boldsymbol{\xi}$:
\begin{equation}\label{scaleimagepos}
	\mathbf{x} = \frac{\boldsymbol{\xi}}{\xi_{0}}.
\end{equation}
For a single thin lens, $\kappa$ is equivalent to the scaled surface density of the lens, and is related to the gravitational potential via the Poisson Equation
\begin{equation}\label{poisson}
\frac{1}{2}\nabla^{2}\psi_{2D} = \kappa = \frac{\Sigma}{\Sigma_{\mathrm{crit}}},
\end{equation}
where the critical surface density for gravitational lensing is given by
\begin{equation}\label{sigmacrit}
\Sigma _{\mathrm{crit}} = \frac{c^{2}}{4\pi G}\frac{D_{s}}{D_{d}D_{ds}}.
\end{equation}
Thus lensing is characterised by $D_{s}$, $D_{d}$ and $D_{ds}$, the angular diameter distances from the observer to the source, from the observer to the (thin) lens, and from the lens to the source respectively.
Total shear can be written in complex notation; its dependency on the two orthogonal components are given by:
\begin{equation}\label{shearcomps}
	\boldsymbol{\gamma} = \gamma_{1} + \mathit{i}\gamma_{2},
\end{equation}
where the two components are linearly related to the second derivatives of the (projected) gravitational potential along two orthogonal directions by
\begin{equation}\label{shearvpot}
	\gamma_{1} = \frac{1}{2} (\psi_{,11}-\psi_{,22})
	 \qquad \mathrm{and} \qquad
	\gamma_{2} = \psi_{,12},
\end{equation}
where the indices after the comma denote partial differentation and we temporarily drop the subscript $2D$ for clarity.
The deformation of the beam is described as a mapping from the source plane to the image (observed) plane. The Jacobian, $\pmb{A}$, of the lens mapping is a real and symmetric $2\times2$ matrix given by
 \begin{equation}\label{lensmap}
 \pmb{A} \equiv 
 \begin{pmatrix}
 1-\kappa - \gamma_{1} & - \gamma_{2}                      \\
  - \gamma_{2}                 &  1-\kappa + \gamma_{1}   \\
 \end{pmatrix}.
 \end{equation}
The flux magnification of an image is given by the inverse of the determinant of the Jacobian, so
\begin{equation}\label{magnification}
	\mu = \frac{1}{(1-\kappa)^{2} - \lvert\gamma\rvert^{2}}.
\end{equation}
One may now consider the effect of lensing on the apparent position of the image of the source of light. 
The scaled deflection angle is the gradient of the lensing potential:
\begin{align}
	\boldsymbol{\alpha_{2D}} & = \nabla\psi_{2D} \label{defangleA}\\
		 	& = \frac{1}{\pi}\int_{\mathbb{R}^{2}} d^2x'\kappa(\mathbf{x}')
			 \frac{\mathbf{x}-\mathbf{x'}}{\left|\mathbf{x}-\mathbf{x'}\right|^2} .\label{defangleB}
\end{align}
The gravitational lens equation is therefore given by
\begin{equation}\label{lenseqn}
	\boldsymbol{\beta} 
	= \boldsymbol{\theta} - \boldsymbol{\alpha_{2D}}(\boldsymbol{\theta}),
\end{equation}
where $\bm{\beta}$ is the angular source position and $\bm{\theta}$ is the angular position of the image on the sky. 
The deflection angle, $\boldsymbol{\hat{\alpha}}$, is related to its scaled counterpart by:
\begin{equation}\label{defangle}
	\boldsymbol{\hat{\alpha_{2D}}}
	= \frac{\xi_{0}D_{s}}{D_{d}D_{ds}} \boldsymbol{\alpha_{2D}}.
\end{equation}

\subsection{The magnification probability distribution} \label{mpdf}
The probability that a source at redshift $z$ would be magnified by an amount within the interval [$\mu$, $\mu+\mathrm{d}\mu$] is  $p(\mu,z)d\mu$. It satisfies :
\begin{equation}\label{normaliseprob}
	 \int^{\infty}_{0} p(\mu,z)d\mu = 1.
\end{equation}
When the $\mu$PDF is convolved with intrinsic luminosity distributions for standard candles, the result describes the observed spread in magnitudes. 
Flux conservation \citep{W76} demands that $p(\mu,z)$ satisfies
\begin{equation}\label{fluxconserve}
	\langle\mu\rangle \equiv \int^{\infty}_{0}\mu p(\mu,z)d\mu = 1.
\end{equation}
There exists a minimum magnification, indeed a minimum convergence, which corresponds to a line of sight that encounters no matter between observer and source; this is also referred to as an empty beam \citep{DR72}.
The distribution function peaks at values below $\mu=1$ and is highly skewed towards high magnifications \citep{HMF00}. For low redshift sources and/or small lensing optical depths, the high magnification tail will exhibit a power law trend $p(\mu)\propto\mu^{-3}$, which results in a formally divergent standard deviation \citep{VO83}. This result was originally derived analytically for a random distribution of compact lenses where only one lens dominates; however, \citet{P93} have noted that for higher optical depths, the slope of tail may become shallower. The precise shape of the distribution depends on the assumed density profile of the lenses \citep{Y08}.
The spread in this distribution increases with source redshift \citep{BL91}. For a fixed source redshift, the shape of the distribution, particularly for low magnifications, depends mostly on the cosmological parameters $\sigma_{8}$ and the vacuum density parameter, $\OmLo$ \citep{P01}.
The high magnification region suffers from low number statistics and since it represents the effects of strong lensing, cannot be probed by ray-bundle methods, which we now discuss.

%---------------------------------------------------------------------------------------------------------------------------------
\section{Ray Tracing}
\label{method}

\subsection{Multiple lens plane methods}\label{multilensplane}
The multiple lens-plane method requires the total three-dimensional mass distribution to be sliced up into contiguous boxes; each box is then projected onto a plane perpendicular to the line-of-sight, usually placed at the centre of the box. In backward ray-tracing methods, light rays are propagated from the observer to the source with deflections {\it only} occurring in successive lens planes. The deflection at a given plane is the result of matter within that plane only. The formalism for the resulting deflection can be written in terms of the multiple lens-plane equation, derived and developed by \citet{BN86}, \citet{SW88} and \citet{J90}; the source position, $\eta$, after deflection by $N$ lens planes:
\begin{equation}
	\boldsymbol{\eta} = \frac{D_{os}}{D_{o1}}\boldsymbol{\xi}_{1} - \sum^{N}_{i=1}D_{is}\boldsymbol{\hat{\alpha}_{i}}(\boldsymbol{\xi_{i}}),
\end{equation}
where $\xi_{i}$ is the position of the ray in the $i^{\mathrm{th}}$ lens-plane; $\xi_{1}$, therefore, is the position of the image. $D_{os}$, $D_{o1}$ and $D_{is}$ are the angular diameter distances from the observer to the source, from the observer to the first lens-plane, and from the $i^{\mathrm{th}}$ lens to the source respectively.
Notice that high density regions in a plane may not necessarily be dense in real space.

%\subsection{The ray bundle method}\label{rbm}
One approach to studying the statistics of gravitational lensing is the ray-bundle method (RBM), developed by \citet{FWM99}, and see also \citet{F99}. Here, the multiple lens-plane method is used to propagate light rays from the observer to a given source plane.
Instead of employing a grid-based technique for calculating magnifications across the source plane, the approach favoured by other backward ray-tracing codes \citep[e.g.][]{JSW00,P01}, the RBM models each individual line of sight as an `infinitesimal bundle' of light rays around a central ray and follows this bundle back to the source plane. An initially circular image is distorted by the time it reaches the source plane as a result of convergence and shear. These quantities can be determined for this specific line of sight as the image-source association is maintained. In the RBM, N-rays in a regular polygon represent a circular image and their positions at the source plane are fitted with an ellipse; the Jacobian matrix (Eqn.~\ref{lensmap}) is determined for each bundle and solved to determine $\mu$, $\kappa$ and $\gamma$. 

The ray-tracing method presented in this work uses the RBM design of a ray-bundle with 8 rays, but does away with the multiple lens-plane treatment of the lensing mass. Instead, the evolution of the cross section of the bundle is determined by integrating the null geodesic using a numerical gravitational potential obtained from N-body simulations.

\subsection{The weak field metric}\label{metric}
The Friedmann-Lema\^{i}tre-Robertson-Walker (FLRW) geometry describes a universe that is homogeneous on all scales
The FLRW line-element for a flat geometry is:
\begin{equation}\label{flrwmet}
ds^2=-dt^2+R^2(t)\left[d\chi^2 + \chi^{2}\left(d\theta^2 + \textrm{sin}^2\theta \, d\phi^2\right)\right],
\end{equation}
where $\chi$, $\theta$ and $\phi$ are comoving coordinates.
The metric allows an evolving scale factor, $R(t)$. From here onwards, we quantify the scale factor relative to its current value, i.e. at $t=0$:
\begin{equation}\label{scalefac}
	a(t) \equiv \frac{R(t)}{R(0)} .
\end{equation}
We assume that the inhomogeneities present in large scale structure are small enough to be represented as a perturbation to the background FLRW metric, and do not falsify the large-scale predictions made under FLRW geometry.
The resulting line element is:
\begin{equation}\label{expandingmetric}
	ds^{2} = -\big[1 + 2\psi(t,\boldsymbol{x}) \big] dt^{2} + 
	a^{2}(t) \big[1 - 2\psi(t,\boldsymbol{x}) \big] d\boldsymbol{x}^{2}
\end{equation}
where $i$ denotes the three spatial dimensions. The weak-field metric is applicable where $\psi(t,\boldsymbol{x}) \ll a(t)$.
The gravitational potential, which is defined with respect to the local perturbation from a smooth background (see Sec.~\ref{phi}), can be decoupled
\begin{equation}\label{phicotophys}
\psi(t,\boldsymbol{x}) = \frac{\phi(\boldsymbol{x})}{a(t)}
\end{equation}
such that $\psi$ is defined in physical units and $\phi$ in co-moving.
The Christoffel symbols for this metric, presented in Appendix~\ref{AppChristoffel}, are dependent, not only on the gradients of the gravitational potential $\phi$, but $\phi$ {\it itself}, which is defined based on the mass perturbation from a smooth background.  
The Geodesic Equations (see Eqn.~\ref{geodesiceqn} below), which are the second order differential equations for the four coordinates, are then constructed.

\subsection{The gravitational potential}\label{phi}
The perturbation field: 
\begin{equation}\label{perturbation}
\delta(t,\boldsymbol{x}) \equiv \frac{\rho(t,\boldsymbol{x}) - \bar{\rho}(t)}{\bar{\rho}(t)}
\end{equation}
relates the local density, $\rho(t,\boldsymbol{x})$, to the mean matter density, $\bar{\rho}(t)$, the latter of which is given by 
\begin{equation}\label{meanmatterdens}
 \bar{\rho}(t) = \OmM\rho_{c}(t).
\end{equation}
The critical density, $\rho_{c}(t)$, is given by
\begin{equation}\label{critdens}
\rho_{c}(t) = \frac{3H^{2}(t)}{8\pi G},
\end{equation}
where $H(t)$ is the Hubble parameter:
\begin{equation}\label{hubbparam}
H(t) = \frac{\dot{R}(t)}{R(t)}.
\end{equation}
The perturbation is related to the gravitational potential $\psi(t,\boldsymbol{x})$, by
\begin{equation}\label{psi}
\psi(t,\boldsymbol{x}) = -\frac{4\pi G}{c^{2}}\bar{\rho}(t)\int_{\mathbb{R}^{3}} d\boldsymbol{x}'^{3} \frac{\delta(t,\boldsymbol{x})}{\left|\mathbf{x}-\mathbf{x'}\right|}.
\end{equation}
The derivatives of the gravitational potential are then
\begin{equation}\label{derivs}
\frac{d\psi(t,\boldsymbol{x})}{d\boldsymbol{x}^{j}} = \frac{4\pi G}{c^{2}}\bar{\rho}(t)\int_{\mathbb{R}^{3}} d\boldsymbol{x}'^{3} \delta(t,\boldsymbol{x})\frac{\mathbf{x}-\mathbf{x'}}{\left|\mathbf{x}-\mathbf{x'}\right|^{3}}.
\end{equation}
One can relate the numerically determined $\phi$ to the gravitational potential via Eqn.~\ref{phicotophys} and derivatives in a similar fashion:
\begin{equation}\label{phiderivscotophys}
\frac{d\psi(t,\boldsymbol{x})}{d\boldsymbol{x}^{j}} = \frac{1}{a(t)} \frac{d\phi(\boldsymbol{x})}{d\boldsymbol{x}^{j}}.
\end{equation}
Constructing the grids that represent the perturbation field and its derivative requires a few steps. Firstly, if the lensing mass is discretized into particles, then the mass within a cube of co-moving side length $L_{box}$ is assigned to the nodes of a regular grid using the Cloud-in-Cell algorithm \citep[CIC;][]{HE88} in three dimensions; the mean matter density is then subtracted off.
A Fast Fourier Transform is applied to convolve the mass distribution with the appropriate kernels to determine the gravitational field and its derivatives according to Eqns.~\ref{psi} and \ref{derivs}, using the popular software package `Fastest Fourier Transform in the West' ({\tt FFTW}\footnote{http://www.fftw.org/}). 
 Depending on the lens, a periodic mass distribution may be implied, or not. If not, the density grid is zero-padded before performing the FFT convolution. 
At many stages during the integration, the local values of the field and its derivatives needs to be evaluated, so an interpolation scheme was written for this purpose; it is described in detail in Appendix~\ref{AppInterp}. 

There are two numerically intensive parts of the ray-tracing method: the first is the set of Fast Fourier Transforms required to calculate the gravitational potential and its derivatives at every grid point on a 3D mesh; the second is the interpolation required at each time-step to determine the values of the same quantities at the exact position of the light ray.

Note also, that the Fourier-grid resolution sets a lower limit to the scale of structure probed, as the mass is smoothed over this grid scale. This is assuming that this scale is reasonably larger than the softening length employed in the N-body simulations used to model the cosmological mass distribution.

\subsection{Three-dimensional ray-tracing}\label{newcode}
Our approach is based on avoiding the approximation of the multiple lens plane method. Instead, the path of a photon is numerically integrated from the Geodesic Equation:
\begin{equation}\label{geodesiceqn}
	\frac{d^{2}x^{\alpha}}{d\lambda^{2}} = - \Gamma^{\alpha}_{\beta\gamma} \frac{dx^{\beta}}{d\lambda} \frac{dx^{\gamma}}{d\lambda}
\end{equation}
Here $ \Gamma^{\alpha}_{\beta\gamma} $ denotes the Christoffel symbols associated with the metric, $x^{i}$ represents any of the four coordinates specified by the superscript $i$, and $\lambda$ is the affine parameter.

The four second-order differential equations  are reduced to eight coupled first-order differential equations, which are integrated with the affine parameter $\lambda$ as the dependent variable. 
A classical fourth order Runge-Kutta integration scheme with fixed step-size was written and used to perform the integration. The subtlety is that it is the {\it $x^{3}$-coordinate}, rather than the affine parameter, that determines the boundaries of the integration. The exact evolution of the affine parameter will be different for each ray-bundle, and is not known in advance. Therefore, the integral is repeated in small steps of $\lambda$, $\Delta\lambda_{\mathrm{RK}}$,until the $x^{3}$ reaches the value required at the source plane $D_{C}(z_{s})$. The step-size is chosen such that the estimated resultant step in $x^{3}$ is a fixed fraction, $f_{\mathrm{RK}}$, of the Fourier-grid resolution: 
\begin{equation}\label{RKstepsize}
\Delta\lambda_{\mathrm{RK}} = \frac{f_{\mathrm{RK}} L_{box}}{n_{\mathrm{FFT}}\dot{x}^{3}_{0}} ,
\end{equation}
where the overdot denotes the derivative with respect to $\lambda$, the extra subscript $0$ denotes values at $t=0$, and $n_{\mathrm{FFT}}$ denotes the number of points along one side of the Fourier-grid (see Sec.~\ref{phi}).
If the ray overshoots the source plane, a simple linear interpolation, using the current and previous positions, is used to find the final source position. The use of a fixed step-size is deemed appropriate as the values of the Christoffel symbols are interpolated from gridded values, as described in Sec.~\ref{phi}. There would be little information gain from step-sizes much smaller than the grid resolution. Nevertheless, the effect of the choice of step-size is analysed, along with other numerical parameters, with results presented in Sections~\ref{schwarzschild} and \ref{fakenfw}.

\subsection{Evolution of the scale factor}\label{evolvescalefac}
For certain cosmologies, the Friedmann Equations
can be solved to find the specific time dependence of the scale factor; for a spatially flat, radiation-free Lemaitre model ($\Omk=0$, $\OmM + \OmL = 1$), with  $\OmL>0$, this dependence is given by :
\begin{equation}\label{scalefac2}
	a(t) = \left[ 
	\frac{\OmMo}{\OmLo} \sinh^{2} \left(
 	\frac{3}{2}\sqrt{\OmLo}H_{0}(t_{0}-t)
 	\right)
 	\right]^{1/3},
\end{equation}
where $t_{0}$ is given by
\begin{equation}\label{lookbacksingularity}
	t_{0} = \frac{2}{3H_{0}\sqrt{\OmLo}} 
	\sinh^{-1} \frac{\OmLo}{\OmMo}.
\end{equation}
Eqns.~\ref{scalefac2} and \ref{lookbacksingularity} were derived from Eqn.~15.36 in \citet{Hobson06}, the solution for a spatially flat, matter-only Lemaitre model, but with $t$ used to denote {\it lookback time} instead. 

As the photon traverses a single lens box, the scale factor will evolve (i.e.\ decrease) although the co-moving scale of the structure does not change appreciably over this time-scale. The value of the scale factor is required to evaluate the Christoffel symbols, so we evolve the lookback time for the photon as well as the spatial coordinates and use this to determine the scale factor at each position throughout the box.

%---------------------------------------------------------------------------------------------------------------------------------

\section{Compact Lens Models}
\label{models}
Here, we present results of ray-tracing in the vicinity of simple lenses. Both of the lens models used are fairly compact, 
and so the thin lens approximation applied in the analytic solution is suitable. The analytic solutions should be read as magnification for the {\it image} at that location, rather than the total magnification of the source. This is an important distinction. The ray-bundle approach only follows the congruence corresponding to a single image, and therefore cannot account for the total magnification of multiply imaged sources; this is one of the reasons why it should not be used to model strong lensing. Yet for these simple lenses, we are still able to test our results against the known magnification of a single image for an otherwise strongly lensed source. 

\subsection{Schwarzschild lenses}\label{schwarzschild}
The Schwarzschild lens, a singular density point, is the simplest lens one may study. The analytic solution for the magnification for an image at any location is well known and is presented in Appendix~\ref{AppSchw}. Since the gravitational potential here is equivalent to a kernel (see Eqn.~\ref{psi}) multiplied by the mass of the lens, we directly compute the gravitational field at each grid point, before performing the ray-tracing. The solution for this lens is scale-invariant, so by increasing the mass of the lens, we may essentially increase the resolution of the grid, and make it possible to study the behaviour of ray-bundles that pass near (or inside) the Einstein radius. In cosmological simulations, the RBM explicitly avoids this as the images near the Einstein ring are part of multiple set of images, and the images inside the ring are the demagnified images which contribute only a small portion of the magnification of the associated source. However, for testing purposes, we allow lines of sight that approach the Einstein radius since the analytic solution is known for each individual image.
In the fiducial test case, the Schwarzschild lens has a mass of $10^{15}\msol$. It is placed at a redshift of $z_{L}=0.35$, giving it an Einstein radius of $65.\mspace{-5.0mu}"6$ for a source redshift of $z_{s}=0.8$, which corresponds to $\sim12$ mesh-points on the $256^{3}$ Fourier-grid.

\begin{figure*}
	\begin{center}  
    	\begin{minipage}[ht]{0.49\textwidth}
	\includegraphics[width=\linewidth]{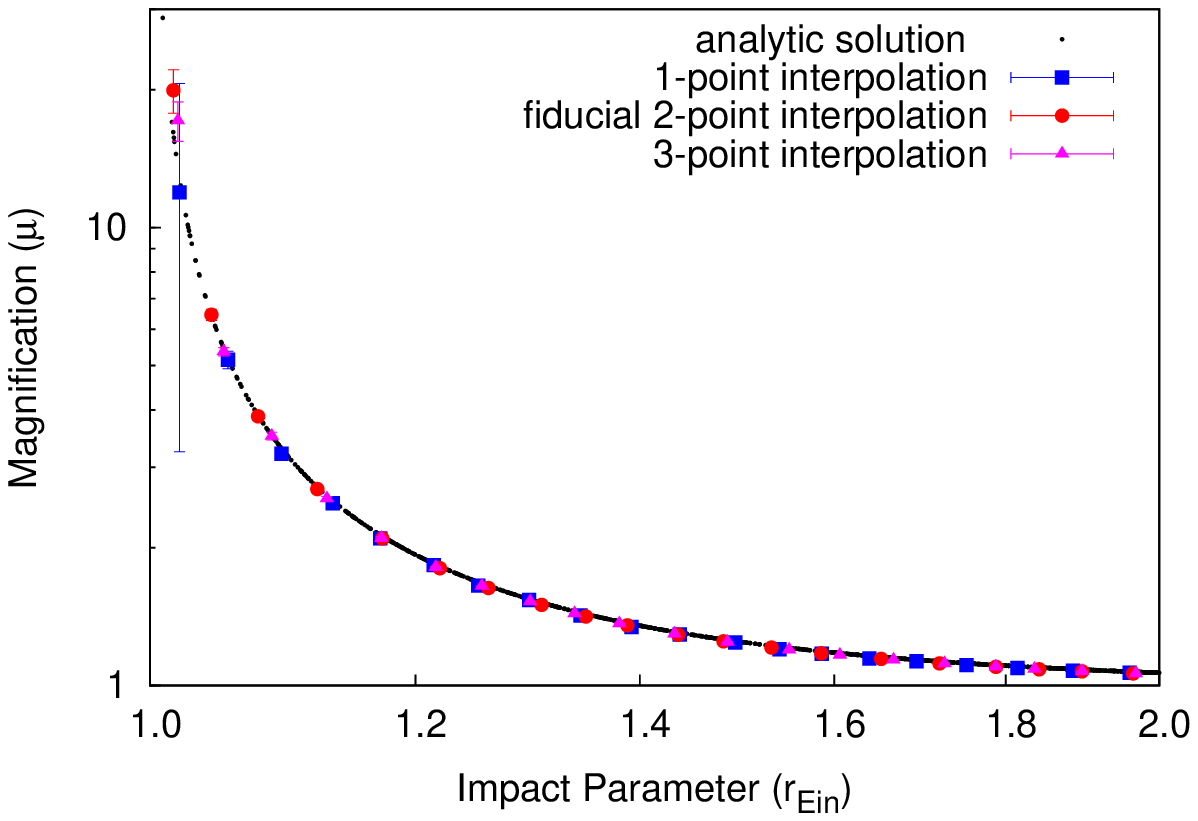}
      	\includegraphics[width=\linewidth]{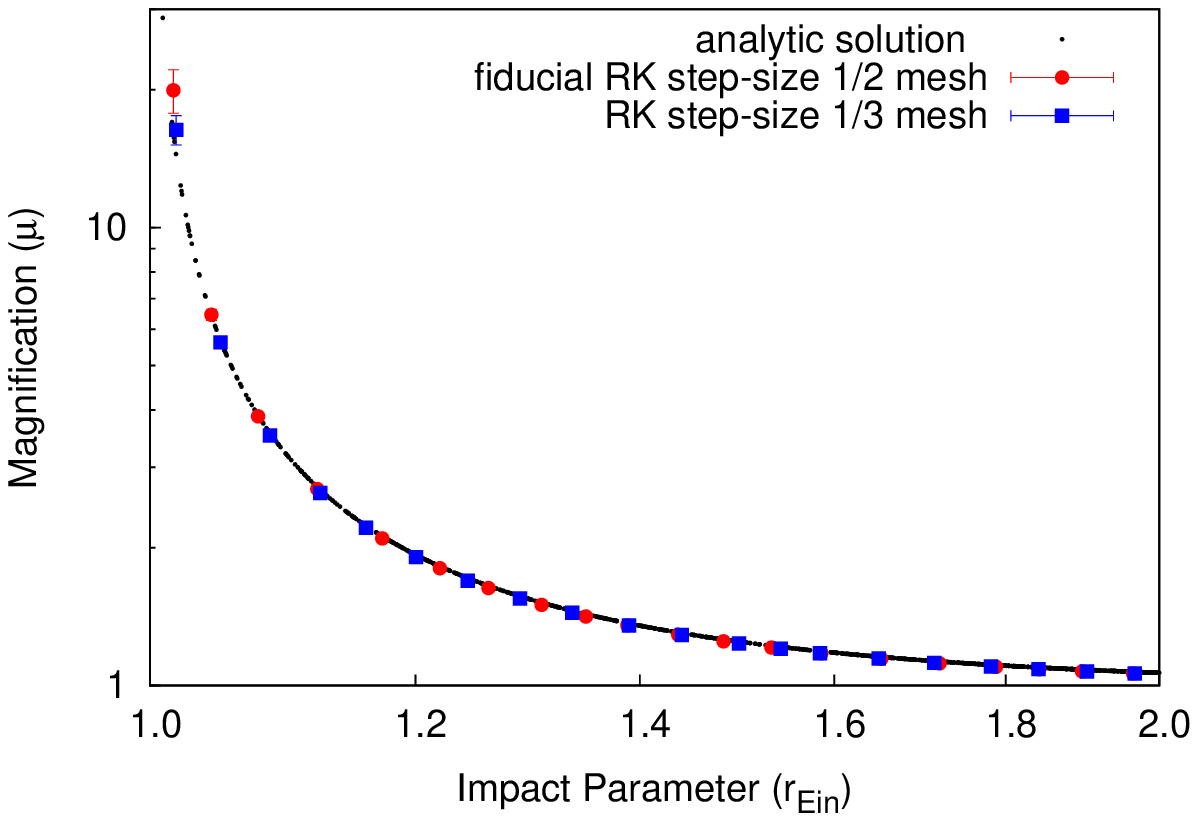}
      	\end{minipage}
     	 \hfill
      	\begin{minipage}[ht]{0.49\textwidth}
      	\includegraphics[width=\linewidth]{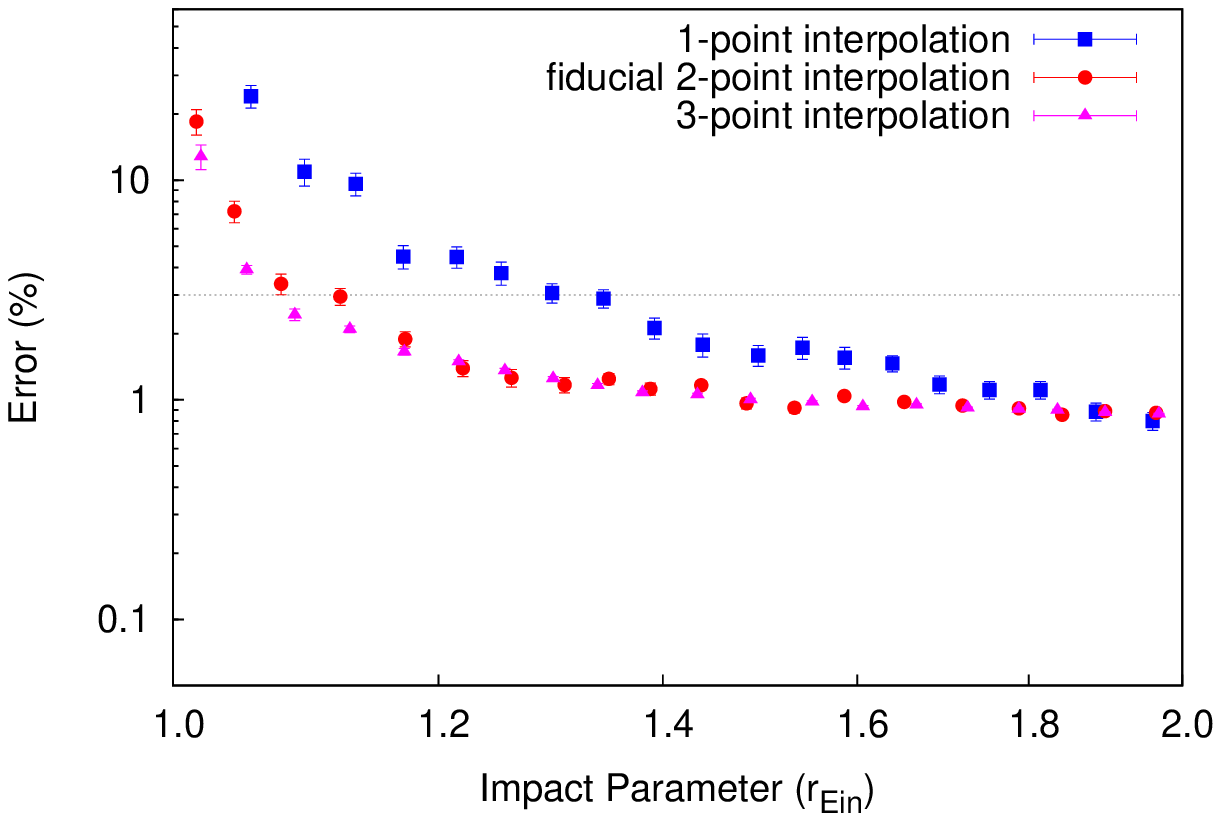}
      	\includegraphics[width=\linewidth]{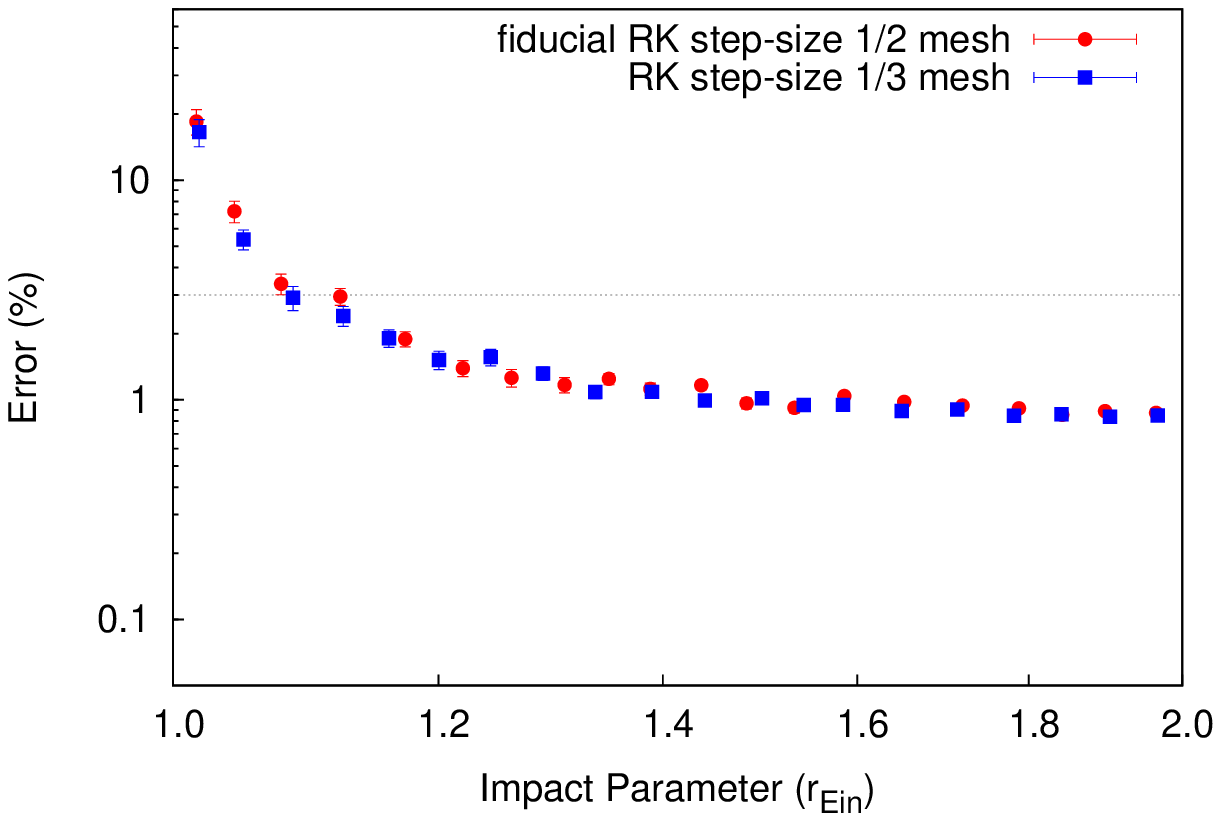}
     	\end{minipage}
      	\caption{The radial dependence of magnification ($\mu$; left) and the percentage error in the numerical magnification ($\Delta\mu/\mu$; right) when ray-tracing past a Schwarzschild lens. 
      Different values of numerical parameters are compared; results are shown for 
      (Top) Number of points used to interpolate the local potential field and its derivatives: $n_{int}=1$ (blue squares),  $n_{int}=2$ (red circles) and $n_{int}=3$ (pink triangles);
      (Bottom) Runge-Kutta step-size that is: 1/2 the Fourier-grid resolution (red circles) or 1/3 the grid resolution (blue squares). 
      In each case, $10^{3}$ lines of sight have been distributed and binned uniformly with log($r/r_{\mathrm{vir}}$) and 1$\sigma$ error-bars are shown. The analytic solution (black dots) at each impact parameter sampled is included in the panels on the left. In the panels on the right, the dotted horizontal line marks the 3 per cent error threshold.}\label{schwtests}
	\end{center}
	\vspace{3mm}
\end{figure*}

The ray-tracing code developed here is able to reproduce the desired lensing distortion very well. The plots on the left hand side of Fig.~\ref{schwtests} show that even high magnification events, which occur when the image is near the Einstein radius, are recovered by the numerical method. 
The numerical error, shown on the right hand side, increases for images that are closer to this radius where the higher-order lensing effects, like flexion, are expected to play a role and an eight-ray bundle has limited application regardless of the choices of other numerical parameters. Note that in each of the tests shown in the Figures~\ref{schwtests} -- \ref{nfwtests3}, the results break down in the very centre of the lens. We choose 3 per cent as an error  threshold to mark the sharp upturn in the average error at small impact parameters. The error surpasses 3 per cent when the impact parameter is smaller than a limiting radius, hereafter referred to as the `minimum reliable radius', or MRR. The MRR can be used as a measure of numerical accuracy and the effect of numerical parameters. For example, the top row of Fig.~\ref{schwtests} shows the result of decreasing the number of interpolation points so that just one Fourier-grid node {\it on either side} of the current position is used to find the local values of the gravitational field and its derivatives. In this case, the MRR expands out to a larger impact parameter where only low magnification regions ($\mu\lesssim1.5$) are recovered with $<3$ per cent accuracy. However, once the number of interpolation points is increased beyond this, this parameter has little influence on the accuracy of the results. Likewise, on the bottom row of Fig.~\ref{schwtests} we show the result of reducing the Runge-Kutta step-size; a step that is approximately half the size of the Fourier-grid resolution ($f_{\mathrm{RK}}=0.5$) is sufficient for accuracy at moderate magnifications ($1.5\lesssim\mu\lesssim6$). Decreasing this parameter has negligible effect on the results.

\subsection{NFW lenses}\label{fakenfw}
Various studies of cosmological simulations \citep{NFW95} have found that dark matter haloes on galactic and cluster scales have mass distributions that are well described by a Navarro-Frenk-White (NFW) profile:
\begin{equation}\label{nfwprof}
	\rho(r) = \frac{\delta_{c}\rho_{c}}{(r/r_{s})(1+r/r_{s})^{2}}.
\end{equation}
Here, $\rho_{c}$ is the critical density (see Eqn.~\ref{critdens}) {\it at the halo redshift}; $r_{200}$ is the radius within which the mean density of the halo is $200\rho_{c}$; $r_{s}=r_{200}/\conc$ is the characteristic scale radius, which marks the transition in the the slope of the profile; and $\conc$ is the dimensionless concentration parameter. Finally, the characteristic overdensity is given by:
\begin{equation}\label{overdens}
	\delta_{c} = \frac{200}{3}\frac{\conc^{3}}{\ln(1+\conc)-\conc/(1+\conc)}.
\end{equation}
We use this profile to compare how the multiple-lens plane approach and the three-dimensional approach model lensing around galaxy and cluster haloes, where most of the high magnification events will occur. 

\begin{figure*}
	\begin{center}  
      \begin{minipage}[ht]{0.49\textwidth}
      \includegraphics[width=\linewidth]{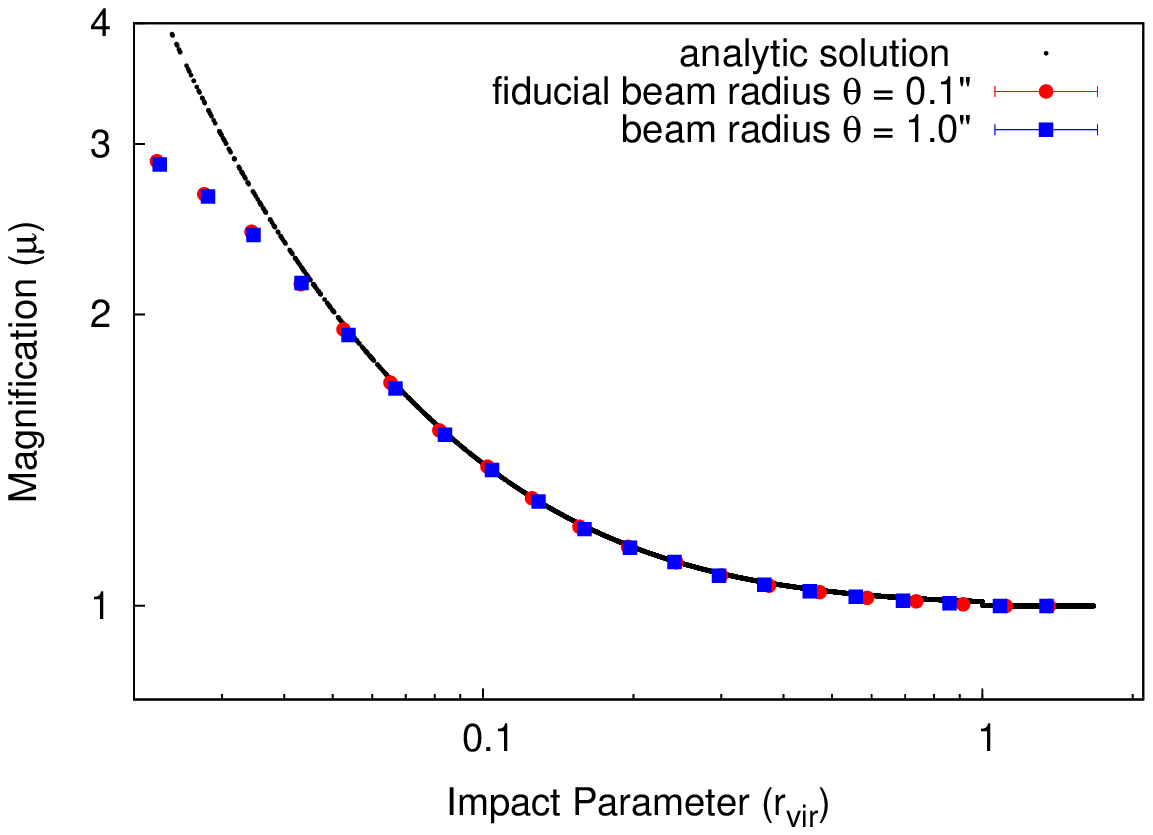}
      \includegraphics[width=\linewidth]{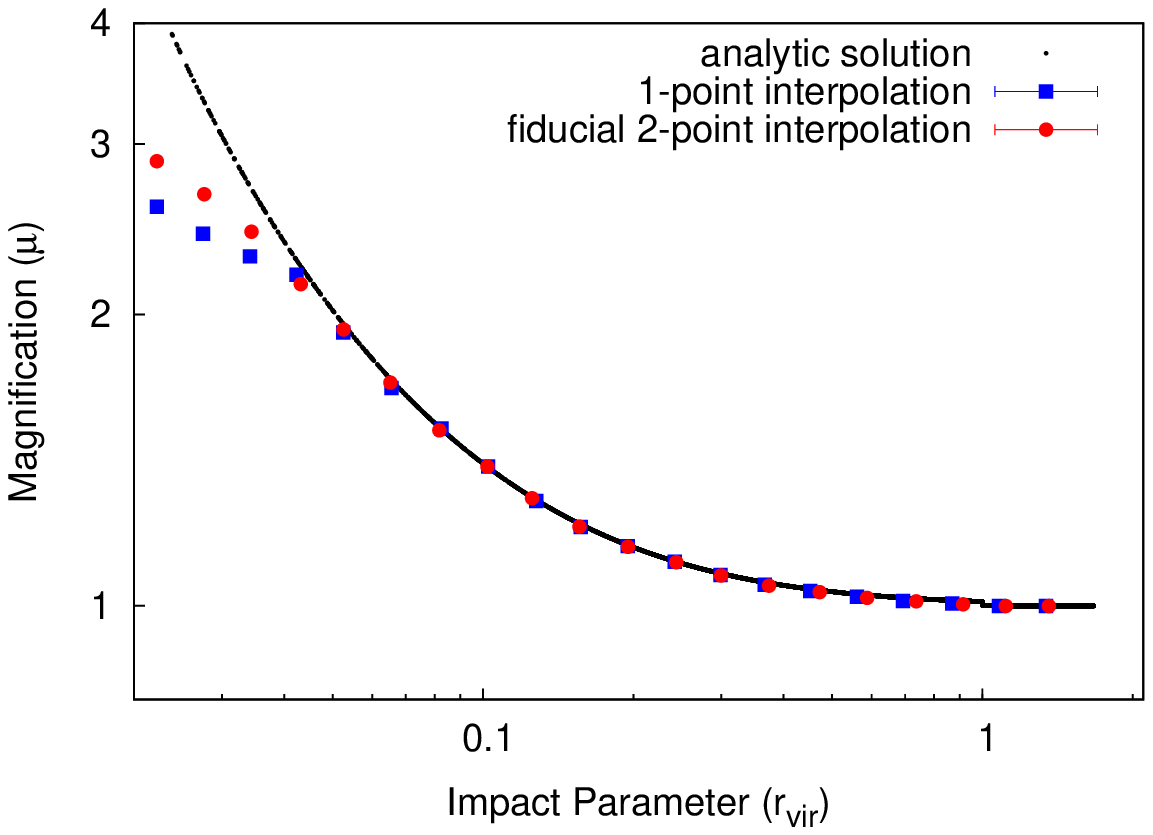}
      \end{minipage}
      \hfill
      \begin{minipage}[ht]{0.49\textwidth}
      \includegraphics[width=\linewidth]{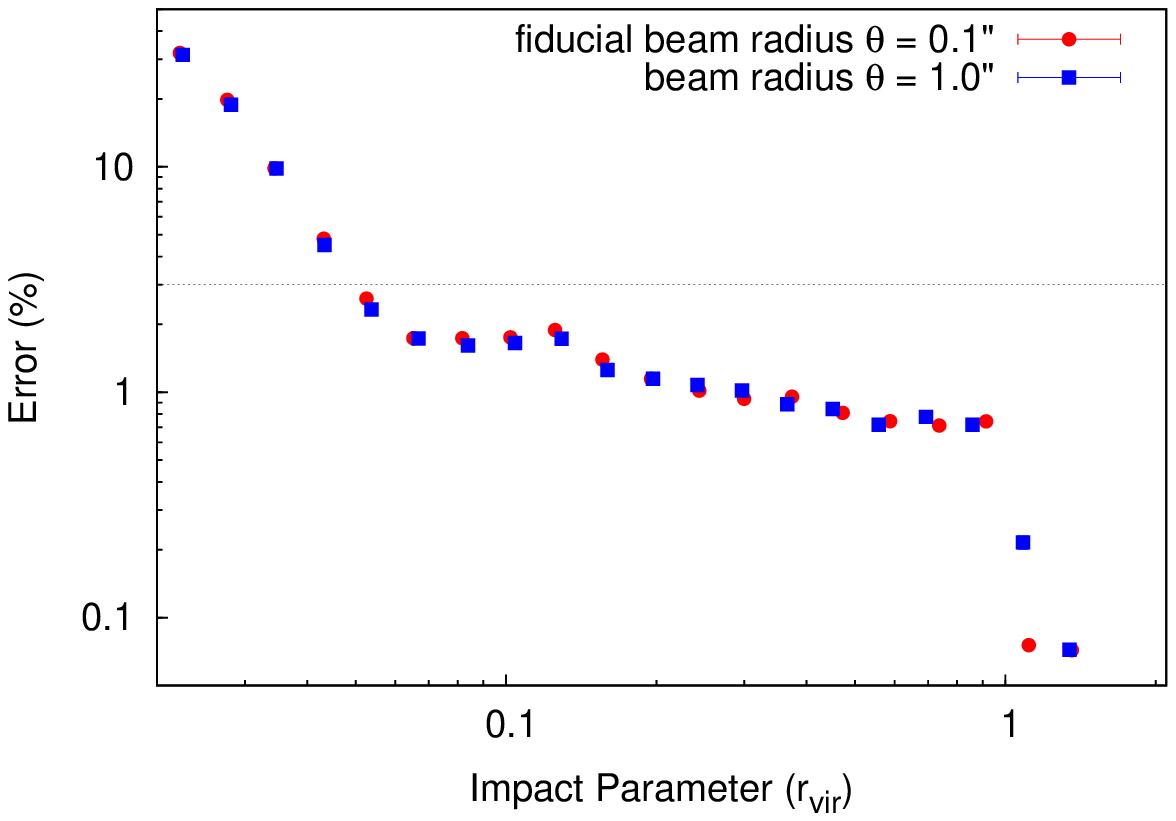}
      \includegraphics[width=\linewidth]{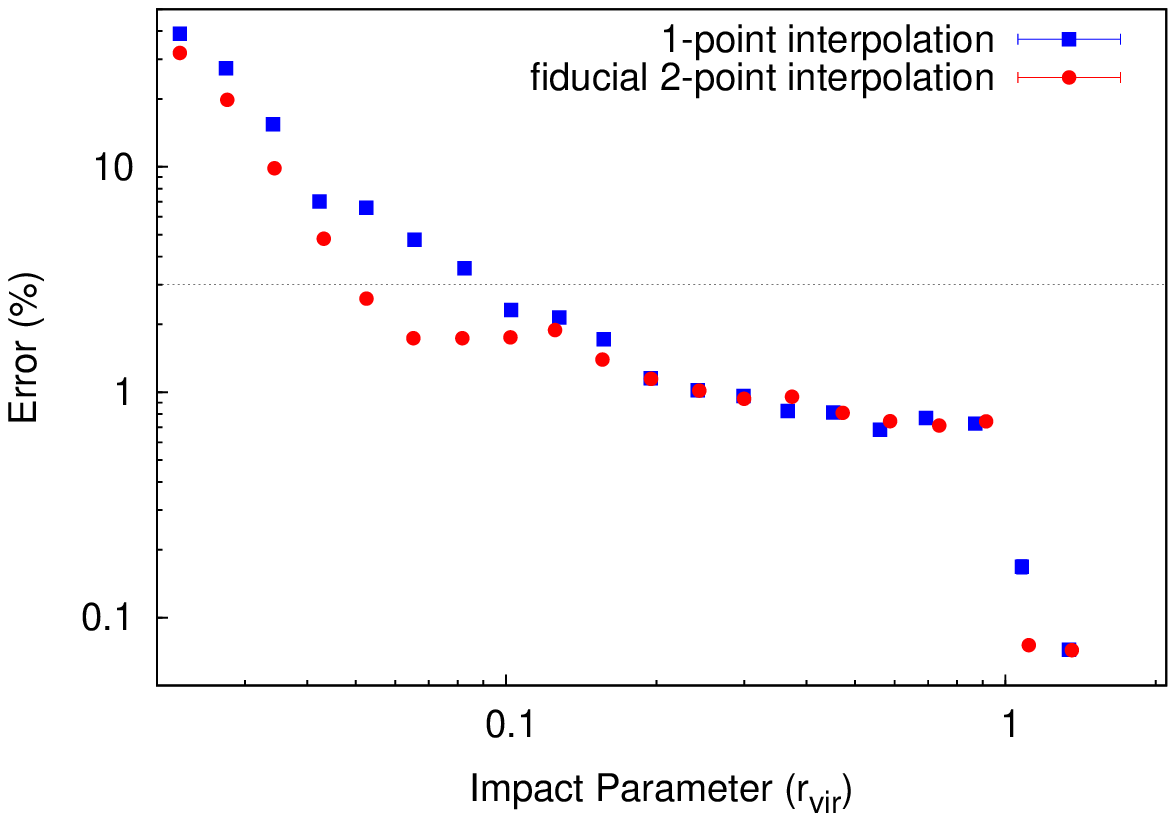}
      \end{minipage}
      \caption{The radial dependence of magnification ($\mu$; left) and the percentage error in the numerical magnification ($\Delta\mu/\mu$; right) when ray-tracing through a discretized NFW lens. 
      Different values of numerical parameters are compared.
      (Top) Radius of ray-bundle/image: $\theta=0.1"$ (red circles) and $\theta=1"$ (blue squares);
      (Bottom) Number of points used in the interpolation scheme: $n_{int}=2$ (red circles) and $n_{int}=1$ (blue squares);
      In each case, $10^{4}$ lines of sight have been distributed and binned uniformly with log($r/r_{\mathrm{vir}}$), and 1$\sigma$ error-bars are shown. The analytic solution (black dots) at each impact parameter sampled is included in the panels on the left. In the panels on the right, the dotted horizontal line marks the 3 per cent error threshold.}
      \label{nfwtests1}
	\end{center}
	\vspace{3mm}
\end{figure*}

\begin{figure*}
	\begin{center}  
      \begin{minipage}[ht]{0.49\textwidth}
      \includegraphics[width=\linewidth]{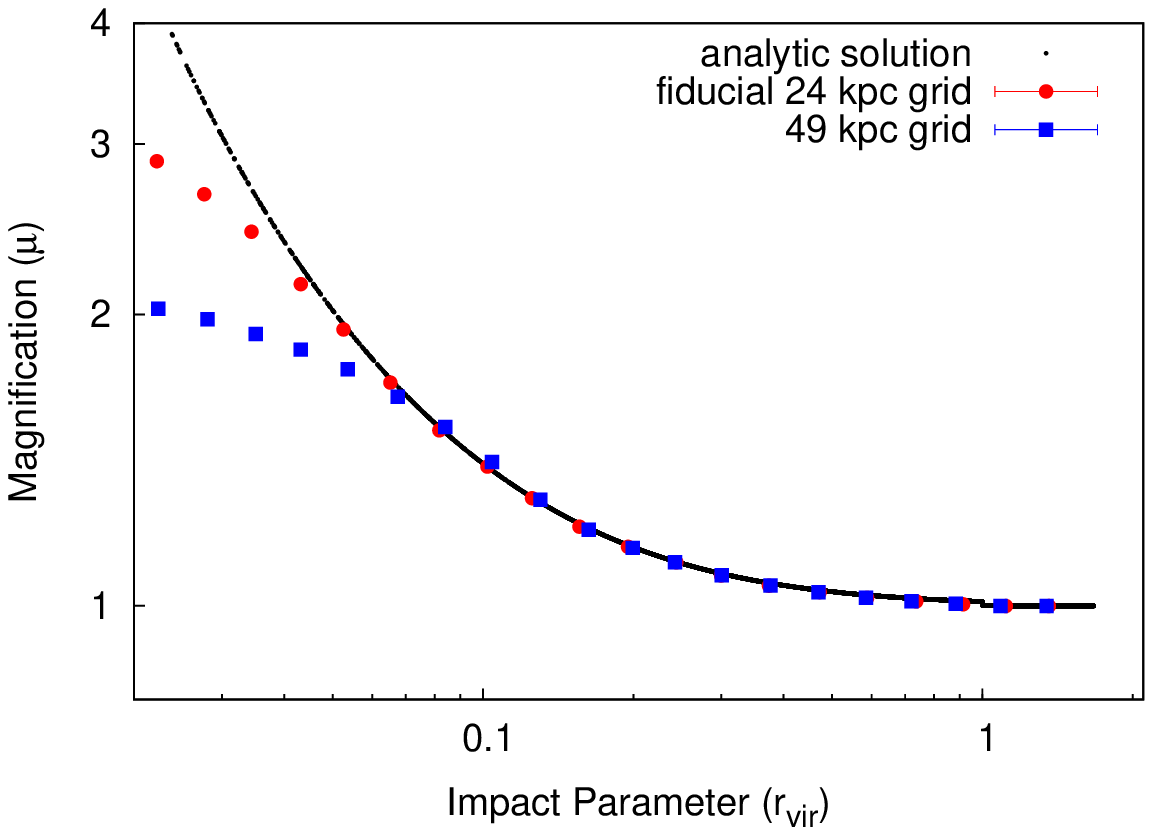}
      \includegraphics[width=\linewidth]{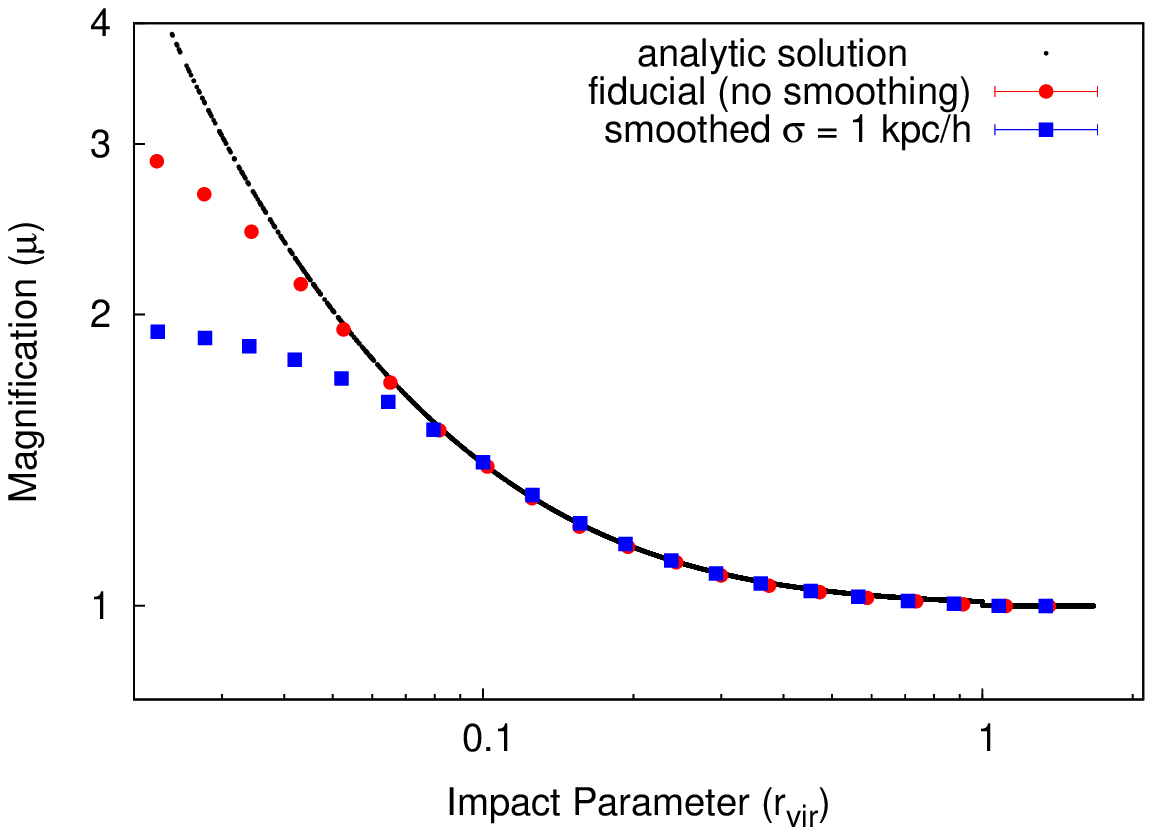}
      \end{minipage}
      \hfill
      \begin{minipage}[ht]{0.49\textwidth}
      \includegraphics[width=\linewidth]{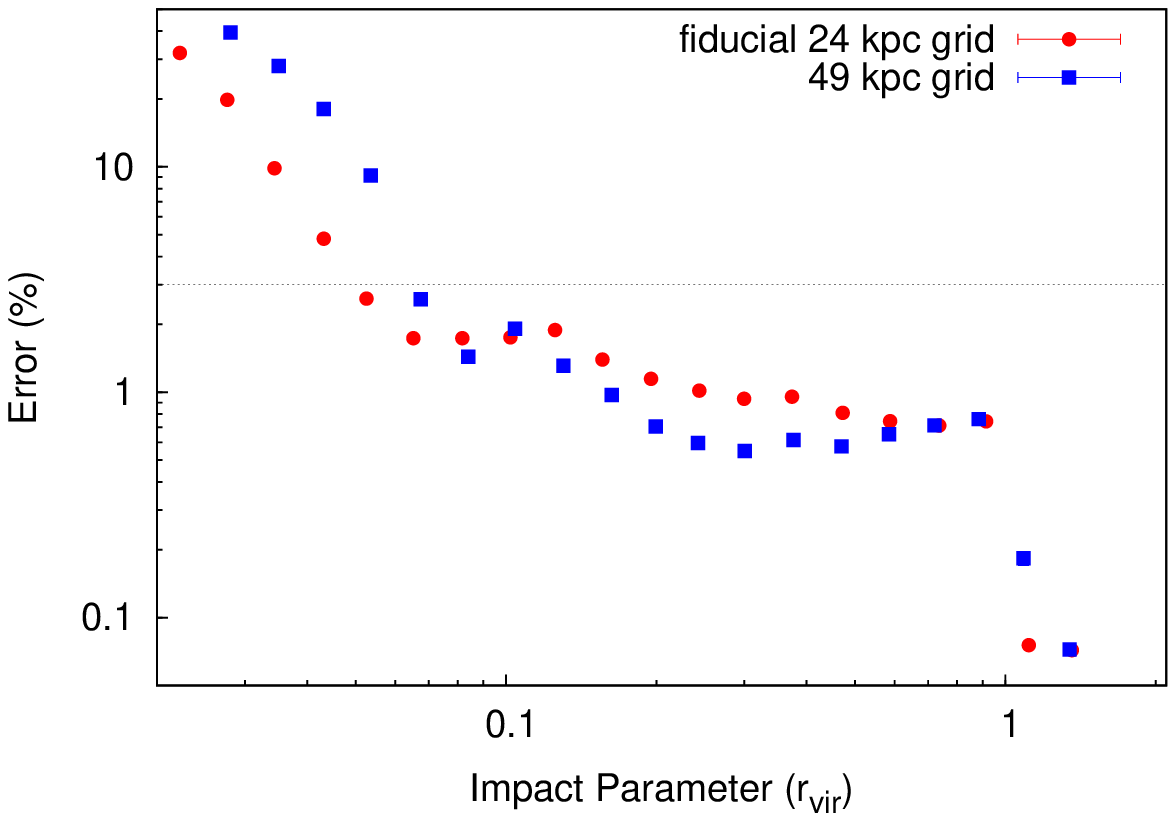}
      \includegraphics[width=\linewidth]{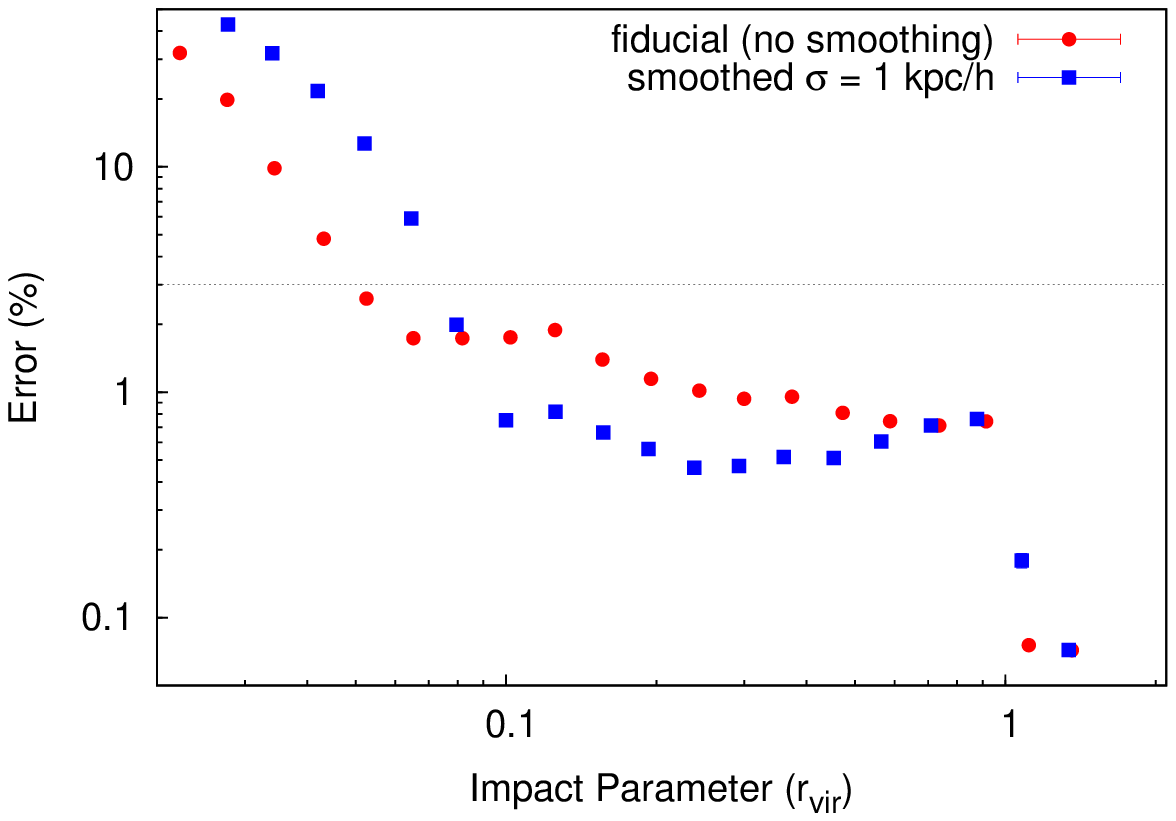}
      \end{minipage}
      \caption{The radial dependence of magnification ($\mu$; left) and the percentage error in the numerical magnification ($\Delta\mu/\mu$; right) when ray-tracing through a discretized NFW lens. 
      Different values of numerical parameters are compared.
      (Top) Fourier-grid resolution: $24$ h$^{-1}$kpc (red circles) and $49$ h$^{-1}$kpc (blue squares).
      (Bottom) No smoothing (red circles) and smoothing with a filter of width $\sigma = 1$ h$^{-1}$kpc (blue squares). 
      In each case, $10^{4}$ lines of sight have been distributed and binned uniformly with log($r/r_{\mathrm{vir}}$), and 1$\sigma$ error-bars are shown. The analytic solution (black dots) at each impact parameter sampled is included in the panels on the left. In the panels on the right, the dotted horizontal line marks the 3 per cent error threshold.}
      \label{nfwtests2}
	\end{center}
	\vspace{3mm}
\end{figure*}

\begin{figure*}
	\begin{center}  
      \begin{minipage}[ht]{0.49\textwidth}
      \includegraphics[width=\linewidth]{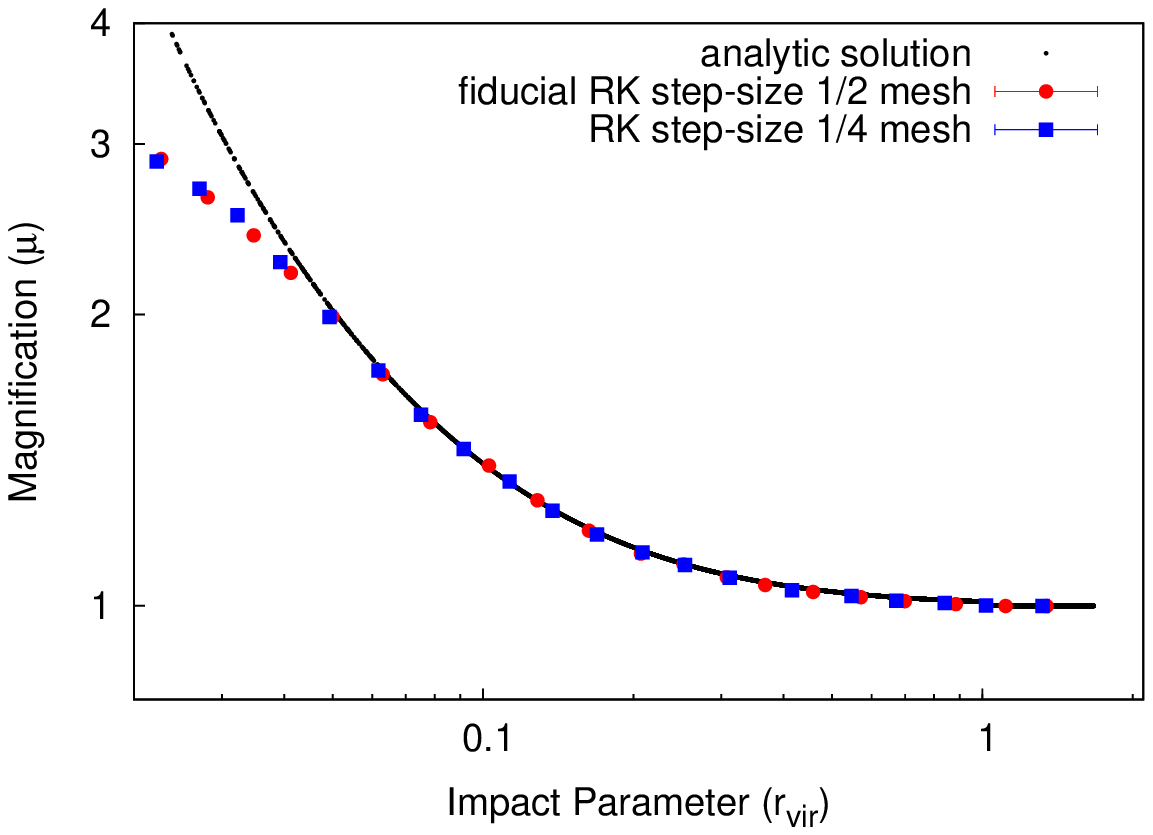}
      \includegraphics[width=\linewidth]{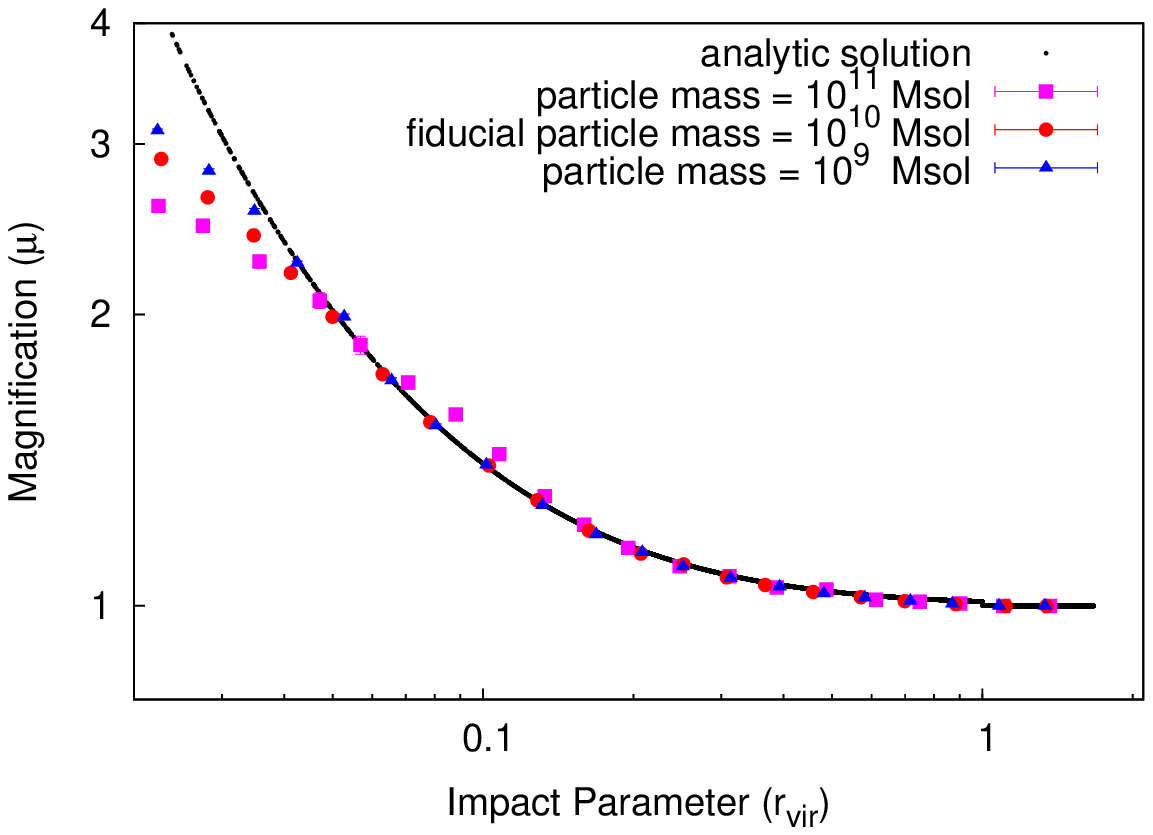}
      \includegraphics[width=\linewidth]{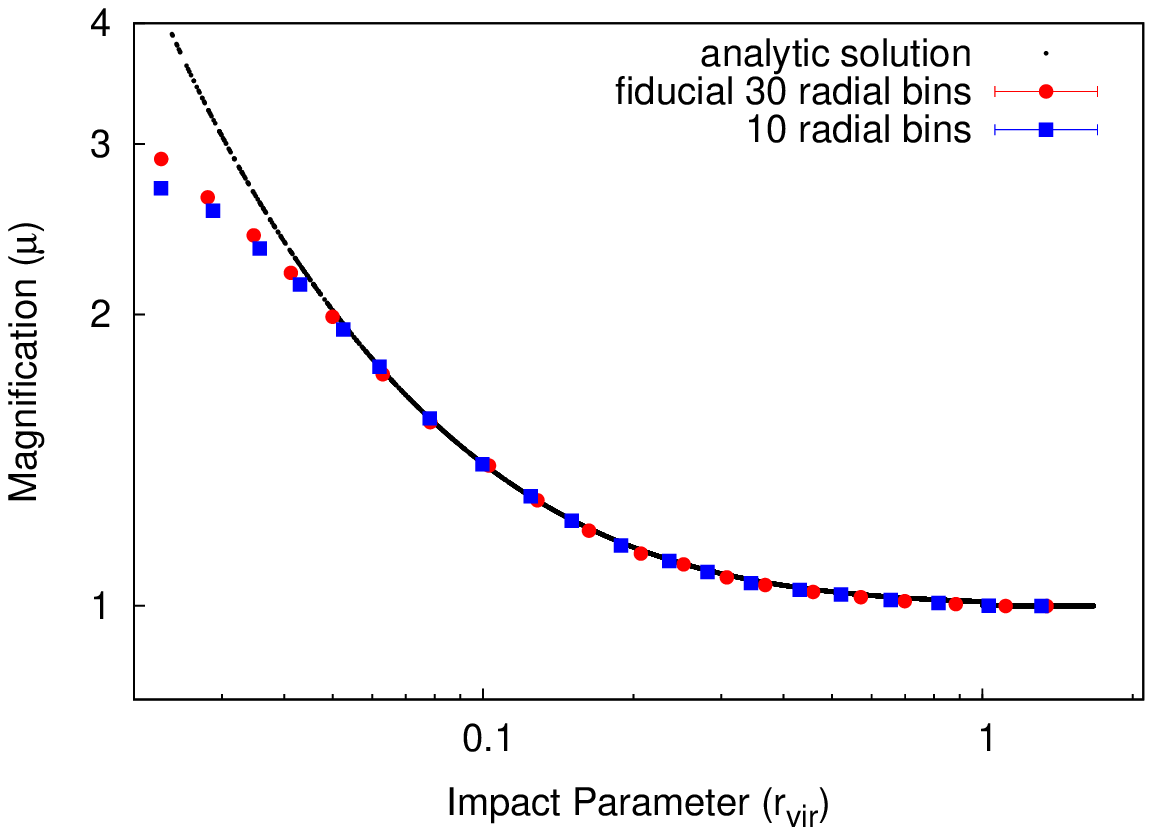}
      \end{minipage}
      \hfill
      \begin{minipage}[ht]{0.49\textwidth}
      \includegraphics[width=\linewidth]{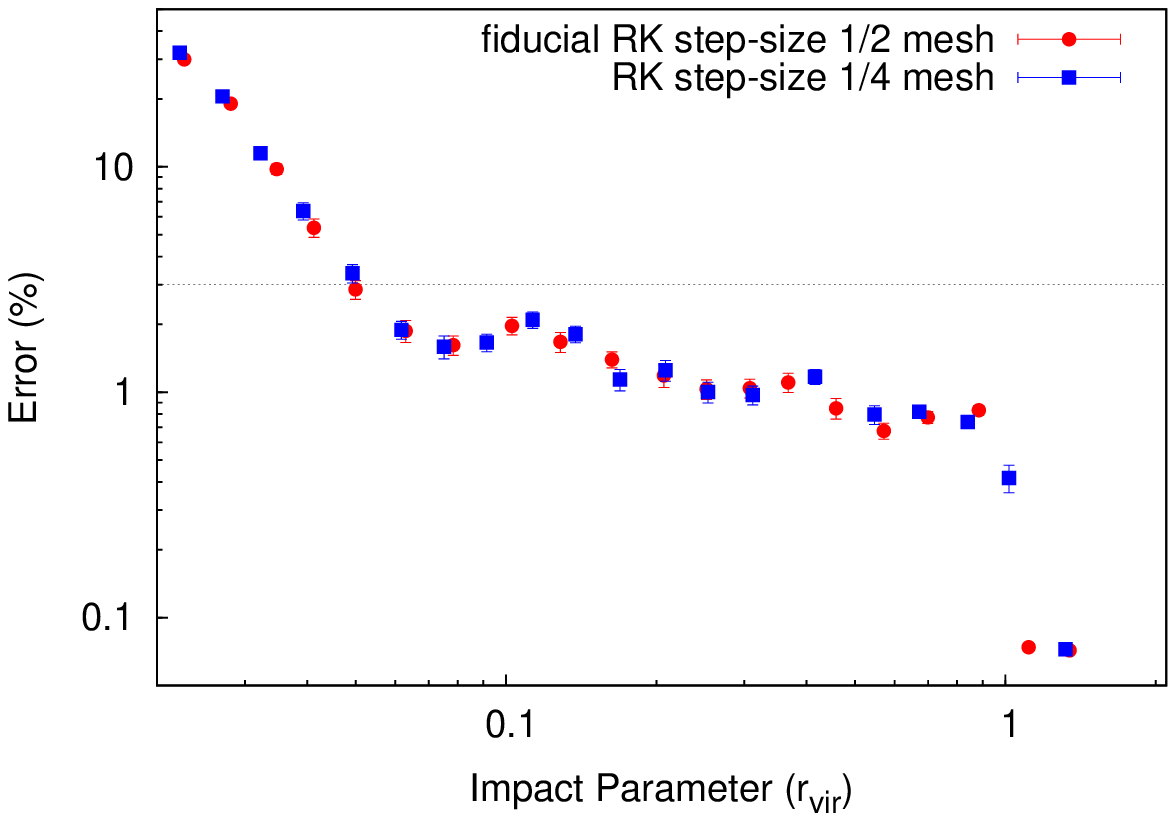}
      \includegraphics[width=\linewidth]{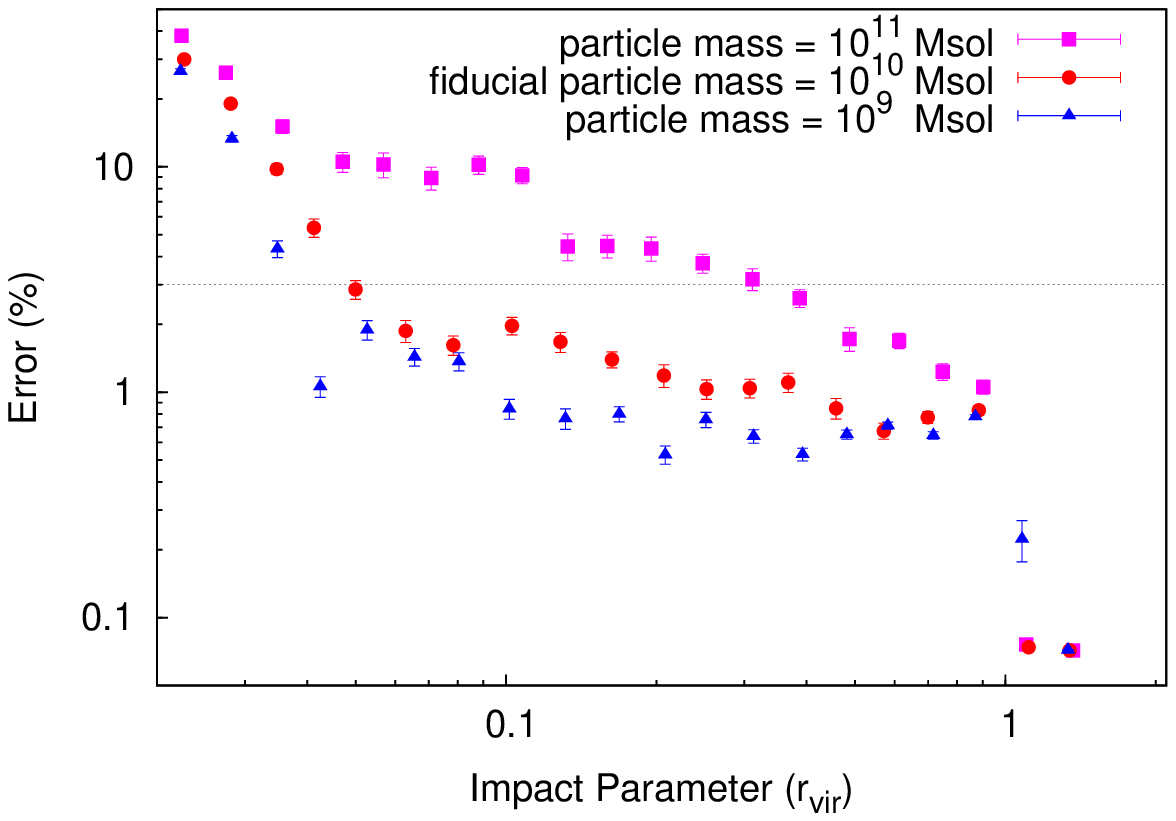}
      \includegraphics[width=\linewidth]{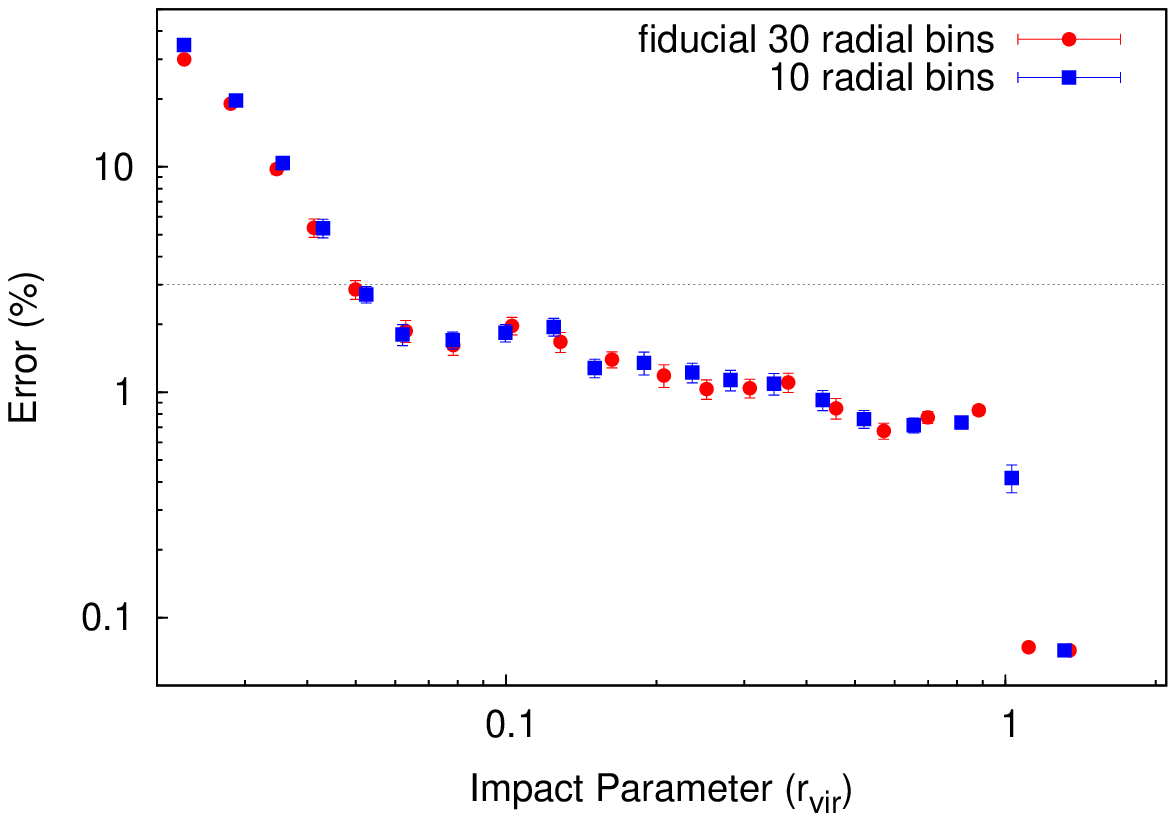}
      \end{minipage}
      \caption{The radial dependence of magnification ($\mu$; left) and the percentage error in the numerical magnification ($\Delta\mu/\mu$; right) when ray-tracing through a discretized NFW lens. 
      Different values of numerical parameters are compared.
      (Top) Runge-Kutta step-size that is: 1/2 the Fourier-grid resolution (red circles) and for 1/4 the grid resolution (blue squares);
      (Middle) Mass resolution: $m_{\mathrm{p}}/\msol=10^{10}$ (red circles), $m_{\mathrm{p}}/\msol=10^{11}$ (pink triangles) and $m_{\mathrm{p}}/\msol=10^{9}$ (blue squares); 
      (Bottom) Number of radial bins used for the lens discretisation: 30 (red circles) and 10 (blue squares).
      In each case, $10^{3}$ lines of sight have been distributed and binned uniformly with log($r/r_{\mathrm{vir}}$), and 1$\sigma$ error-bars are shown. The analytic solution (black dots) at each impact parameter sampled is included in the panels on the left. In the panels on the right, the dotted horizontal line marks the 3 per cent error threshold.}
      \label{nfwtests3}
	\end{center}
	\vspace{3mm}
\end{figure*}

Lenses with the NFW profile were modelled by discretising the mass contained in one virial radius into a number of particles of equal masses and constructing a fake simulation output in the {\tt GADGET} format. The numerically determined gravitational potential is therefore {\it not} equivalent to the analytic solution, but suffers from discretisation effects just as a simulated halo would. 
The spherically symmetric lens is divided into a fixed number of radial bins of equal width, extending from the centre to the virial radius. Each particle is randomly placed in one of the radial bins with a probability proportional to the mass within that bin; the bin mass is found by appropriately integrating the density profile, given in Eqn.~\ref{nfwprof}. The lenses modelled for this purpose have a virial mass, and total mass, of $M_{vir}/\msol = 10^{14}$ and concentration parameter $\conc=7.2$. At a (lens) redshift of $z_{L}=0.35$ and a virial overdensity of $\Delta_{c}= 200$, their virial radii are $0.55$ h$^{-1}$Mpc. The results presented here are for a source redshift of $z_{S}=0.8$.  
Figures \ref{nfwtests1} -- \ref{nfwtests3} show the magnifications determined by ray-tracing over a range of impact parameters changing one numerical parameter relative to a fiducial choice. The ray-bundles are allowed to encounter the lens at a distance of up to twice the virial radius, which corresponds to an angular separation of $343"$. 
The analytic solutions for image magnification and shear by a lens with the NFW profile are presented in Appendix~\ref{AppNFW}. Since the model lens is truncated at one virial radius, the metric on the exterior is given by the Schwarzschild metric, by Birkhoff's theorem. Thus the analytic solution shown for $\mu(r)$ switches to that appropriate for a Schwarzschild lens outside this impact parameter. 

Between the MRR and the truncation radius, the error remains fairly constant at 1 per cent. Outside the truncation radius, the error drops sharply by an order of magnitude, which implies that the removal of mass outside this radius is primarily responsible for the aforementioned 1 per cent error.
The size of the ray-bundle, or equivalently, the radius of the image or beam, has negligible effect on the ray-tracing results (see top row of Fig.~\ref{nfwtests1}). 
By varying the grid size (see top row of Fig.~\ref{nfwtests2}), we notice that the MRR has a very strong dependence on the Fourier-grid resolution. Our interpretation is that the interpolation scheme used to measure the local gravitational field would smooth over the central cusp, causing the large errors here. We conclude that if small-scale structure is responsible for a caustic (high magnification) then this will be washed out.
The MRR has a strong dependence on the number of points used to interpolate the values of the gravitational potential and its derivatives from the Fourier-grid on which they are calculated, as shown in the bottom row of Fig.~\ref{nfwtests1}. 
The use of more interpolation points allows for more robust ray-tracing, with a more accurate mean magnification value.
However, the computation time rises approximately linearly with the number of interpolation points used, and the interpolation scheme suffers in the vicinity of density peaks. The latter issue should not pose too much of a problem when the lensing mass distribution is large-scale structure, but computational efficiency presents a significant hurdle for ray-tracing procedures such as this.
The parameters that most govern the accuracy of the ray-tracing scheme are the Fourier-grid size and mass smoothing length, which have essentially equivalent effects; assigning mass to an Fourier-grid acts to spread it out over a fixed number of grid-points. A CIC mass assigning scheme is different to a Gaussian filter smoothing, but in essence, the larger the area over which mass is spread, the more density peaks are suppressed and the larger the error in the tests shown in Fig.~\ref{nfwtests2}. On the top row of Fig.~\ref{nfwtests3}, we show that halving the Runge-Kutta step-size has a negligible effect, so half the Fourier-grid resolution is deemed sufficient. 
The model lens used for this test is subject to a few user-defined variables; a few convergence tests were run to ensure that these were not the cause of the observed features. Trialling different mass resolutions from $10^{9}$ to $10^{11} \msol$ for a fixed halo virial mass, $M=10^{14}\msol$, the results are shown in the panels of the middle row of Fig.~\ref{nfwtests3}. We find that the resulting scatter increases from less than 1 per cent up to 10 per cent at a fixed impact parameter. One could equivalently say that the MRR is larger for haloes that are not as well resolved. The cosmological mass distributions in Sec.~\ref{sims} have mass resolutions better than $10^{10}\msol$, so one can expect a low error in magnification even within dense regions.
The number of bins that the lens is divided into is a choice that has negligible effect on the ray-tracing results, as shown in the bottom row of Fig.~\ref{nfwtests3}.
One is therefore reassured that the results of the ray-tracing on the discretised NFW profile are relevant to the results of ray-tracing through cosmological mass distributions based on the fiducial parameter choices.
\citet{KF08} numerically integrate the null geodesic equations (with the adaptive stepsize Runge-Kutta-Fehlberg 4-5 method) in order to test the accuracy of the thin-lens approximation, but only with reference to strong lensing by singular isothermal sphere (SIS) and Navarro-Frenk-White \citep[NFW;][]{NFW95} mass profiles, both of which are relatively thin.

%---------------------------------------------------------------------------------------------------------------------------------

\section{Large-Scale Structure}
\label{results}

\subsection{Cosmological simulations}
\label{sims}
In order to construct the predicted probability distributions for the weak lensing statistics numerically, numerous null geodesic equations are integrated through a cosmological mass distribution that is generated with N-body simulations. 
The cosmological simulations are carried out with the parallel Tree-PM SPH code \texttt{GADGET2} \citep{S05} using collisionless particles only. We adopt a \lcdm~cosmology, with the following values for the cosmological parameters:  $\OmMo$ = 0.27, $\OmLo$ = 0.73, $h$=0.71, $\sigma_{8}=0.9$. The dark matter distribution is discretised into $256^{3}$ particles distributed within a periodic box with co-moving length of $L_{\mathrm{box}}=50$~h$^{-1}$Mpc, resulting in a mass resolution of $m_{p}=6.3\times10^{8}\msol$; the initial displacements are in a `glass' configuration. The simulations begin at an initial redshift of $z_{i}=39$, with a redshift-dependent gravitational force softening length of $\epsilon_{co}=16$ h$^{-1}$kpc (Plummer-equivalent). 

The space between an observer and source is divided up into individual regions, each modelled by a snapshot of a cosmological simulation at an appropriate redshift. The snapshot cadence, $\Delta z$, is chosen such that the light travel time corresponds to the length of the boxes. The line-of-sight integrated co-moving distance between an observer and source at $z=z_{s}$ is:
\begin{equation}\label{raytracedistance}
D_{c} = c\int_{0}^{z_{s}}\frac{dz}{H(z)}.
\end{equation}
Since this cannot be easily evaluated in general, we instead assume that the differences between the redshift of adjacent boxes are small, and thus:
\begin{equation}\label{boxsizestack}
L_{\mathrm{box}} = \frac{c \Delta z}{H(z)},
\end{equation}
which is easily rearranged to determine the snapshot cadence. A total of 28 boxes are required to fill the space between an observer and a source at $z\approx0.5$.

Note that the scale factor at the position of a photon is required to integrate the null geodesic equations. This scale factor is {\it not} identified with the `average' redshift of the current snapshot being traced through; instead it is derived from the lookback time (see Eqn.~\ref{evolvescalefac}) and evolves during the integration through the box. That is to say, it is assumed that in the time it takes for a photon to traverse a box, the co-moving scale of the structure remains constant, however the physical scale length changes. As a sanity check, the numerically derived value of the scale factor at each box interface is compared to the redshift of the appropriate snapshot.

\subsection{The sampling region}
\begin{figure*}
	\begin{center}  
      		\includegraphics[trim=15mm 15mm 15mm 0mm, clip, width=\linewidth]{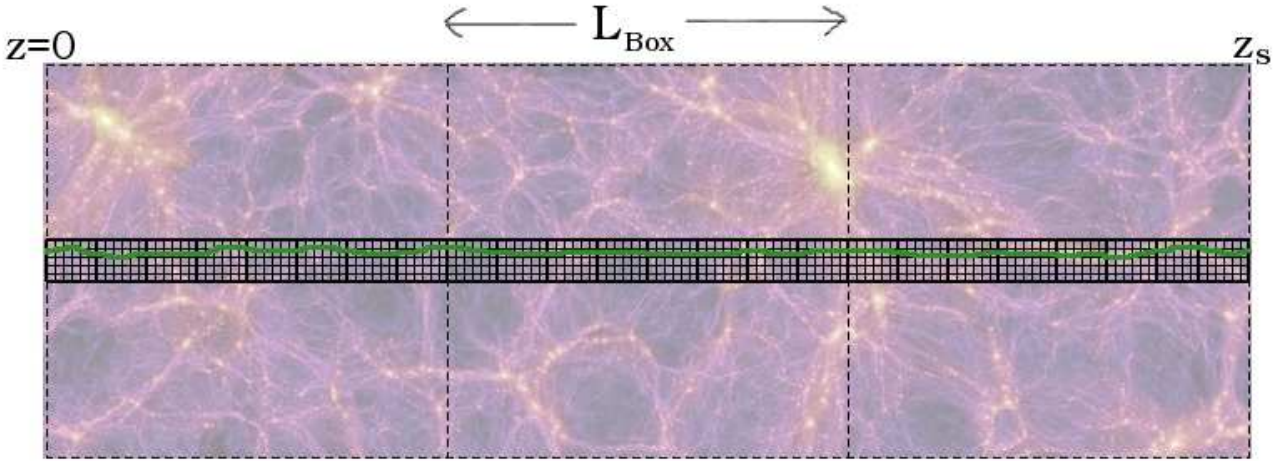}
      \end{center}
	\vspace{1mm}
	\caption{A two-dimensional schematic diagram of the ray-tracing method. Large simulation snapshots (dashed lines; three shown here) of length $L_{\mathrm{box}}$ fill up the space from the observer, $z=0$, to the source plane at $z_{s}$. Small cubes (thick lines) from within these boxes form a prism of lensing mass. The mass in each cube is distributed onto Fourier-grids (thin lines) before calculating the gravitational potential and derivatives in order to numerically integrate the path (green) of a photon.}
	\vspace{5mm}
	\label{RayTracingDiagram}
\end{figure*}

In order to avoid repeated structure, a number of precautions were taken. Nine independent simulations were run, resulting in nine separate realisations. The snapshot for each required redshift was chosen randomly from among the nine realisations. The chosen box was then randomly translated and rotated $90^{\circ}$ about any/all of the axes. From each 50 h$^{-1}$Mpc box we selected the lensing mass formed by particles within a prism 6.25 h$^{-1}$Mpc across the sky and 50 h$^{-1}$Mpc along the line of sight. This is split into 8 cubes each of volume $ V = 6.25^{3}$ (h$^{-1}$Mpc)$^{3} \approx 244$ (h$^{-1}$Mpc)$^{3}$.  Each of these has its mass placed on a $128^{3}$ point grid, using the CIC algorithm. We apply an FFT to these boxes with zero-padding so that the density field (and the derivatives of the gravitational potential) due to structure {\it within the box only} is determined on a scale of 49 h$^{-1}$kpc. Note here that the simulation is run on a larger box such that the large-scale modes are included in the formation of structure, but we only use a smaller portion of the box for the lensing (see Fig.~\ref{RayTracingDiagram} for a cartoon diagram). The null geodesic equations of the eight rays and one anchor of each ray-bundle are numerically integrated using the methods described in Sec.~\ref{newcode}.
Lines of sight that fall too close to the edge of the boxes will be artificially sheared, so assuming that most of the lensing occurs due to mass within 0.5 h$^{-1}$Mpc, we only send out `beams' that will fall within the central $5.25\times5.25$ (h$^{-1}$Mpc)$^{2}$ region at the source redshift. For a source redshift of z=0.5, with an angular diameter distance of approximately 900 h$^{-1}$Mpc, this means that the area of sky sampled is about $0.33^{\circ}\times0.33^{\circ}$ or $5\times10^{-6}$ of the entire sky. To increase the sampling area, this entire procedure is used to trace the paths of 10 000 bundles, then repeated for another 10 000 bundles using another lensing mass distribution sampled from the same set of simulations.

\subsection{Numerically determining the $\mu$PDFs}\label{numericalPDF}
Using ray-tracing procedures to construct a magnification probability distribution is effectively equivalent to calculating the angular diameter distance in different directions out to the same co-moving distance. The distribution of angular diameter distances for a given source redshift has been investigated by \citet{HB97}. In the extremal case of a large beam-size, and therefore, large separations between rays, the angular diameter distances would be the same in all directions and the `distribution' would be a single peak at $D_{fb}(z)$, the average FLRW value, implying homogeneity on those scales. However, if the beam-size is small enough, then the inhomogeneities induce a distribution. 
Each ray-bundle probes a single light of sight. More specifically, it describes the lensing that has affected a single image of a fixed size on the sky. The number of bundles that exhibit a magnification of $\mu$ are thus proportional to the probability that an {\it image} is magnified by $\mu$. However, the statistical quantity of interest is the magnification probability distribution for {\it sources}. Therefore, the number counts that are used to produce the magnification probability histograms presented in Sec.~\ref{compareRBM} are weighted by the area covered by the beam at the source plane, or equivalently, the inverse of the magnification. 
\begin{equation}\label{weighthisto}
	P(\mu)d\mu = \frac{F(\mu,\mu+d\mu)}{\mu}
\end{equation}
where  $F(\mu_{1},\mu_{2})$ is the fraction of all ray-bundles for which $\mu_{1}<\mu<\mu_{2}$.
Similarly the mean and standard deviation of these properties are also weighted.

\subsection{Comparison of the PDFs produced with and without lens planes}\label{compareRBM}
\begin{figure}
	 \begin{center}  
     	 \includegraphics[width=0.95\linewidth]{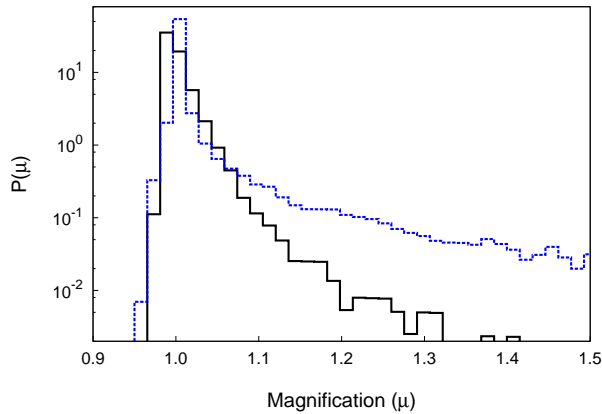}
      	\end{center}
	\caption{The magnification probability distribution for a source at redshift $z=0.5$ as predicted by the multiple lens-plane RBM (blue dashed line) and by the three-dimensional equivalent (black solid line)}
	\label{RBMvsMudMag}
	\vspace{5mm}
\end{figure}

\begin{figure}
	 \begin{center}  
      	\includegraphics[width=0.95\linewidth]{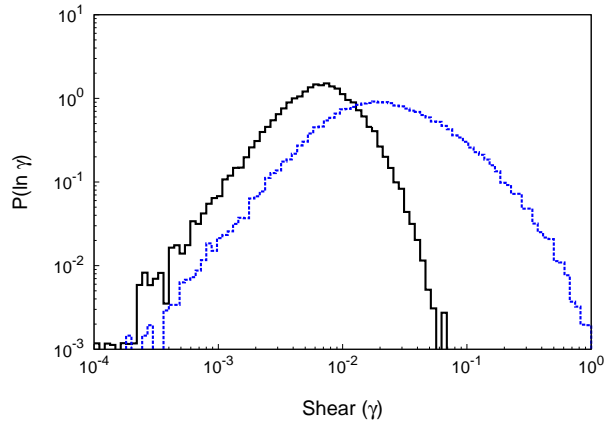}
      	\end{center}
      	\caption{The shear probability distribution for a source at redshift $z=0.5$ as predicted by the multiple lens-plane RBM (blue dashed line) and by the three-dimensional equivalent (black solid line)}
	\label{RBMvsMudShear}
	\vspace{5mm}
\end{figure}

The predicted probability distributions of magnification and shear 
 are constructed by ray-tracing $2\times10^{4}$ bundles through the three-dimensional mass distribution described above. The results, shown in Figures \ref{RBMvsMudMag} and \ref{RBMvsMudShear},  are compared to the probability distributions constructed with the multiple lens-plane RBM using $5\times10^{4}$ bundles. However, not all ray-bundles are included in the analysis. 
The RBM is very well suited to the weak lensing limit and provides a computationally efficient alternative to grid-based ray-shooting methods; however, it underestimates the magnification in cases where multiple imaging is expected as only one image contributes to the magnification of each source. For this reason, rays that fall within a minimum distance of a lens (remembering that a distribution of point masses describes the lensing mass) are excluded from the analysis. This affects the high magnification probabilities deduced, but does not significantly affect the weak lensing analysis \citep{FWM02}. The RBM avoids the danger of artificial shear due to a source pixel collecting light rays that have passed near the edge of a shooting grid.
If a ray-bundle were to suffer from neither convergence nor shear, it would obtain the minimal magnification, $\mu_{eb,min}=1$, or equivalently, $\mu_{fb,min}= D_{fb}^{2}(z_{s})/D_{eb}^{2}(z_{s})$ (see Appendix~\ref{AppBeam}). For a source redshift of $z=0.5$, this minimum magnification is $\approx 0.965$; the relevant angular diameter distances have been determined with the aid of the {\tt Angsiz} routine \citep{KHS97}. Those bundles that produce magnifications below this minimum value represent the demagnified components of a multiple-image system, a possible consequence of strong lensing. These will highly underestimate the total magnification of the associated source, and so, are also excluded in both methods. On the other hand, if a magnified image is traced back to the source plane, but belongs to a multiple-image system, it will not be identified as such. Although it will also underestimate the total magnification, it will generally do so by a negligible amount. The ray-tracing method developed in the present work does not produce any bundles that must be excluded. 
%----------------------------------------------------------------
The average magnification predicted by both methods is close to unity, as required by flux conservation (see Eqn.~\ref{fluxconserve}). More precisely, the multiple-lens plane approach finds $\bar{\mu}=1.0199$ while the three-dimensional approach finds a mean magnification of $\bar{\mu}=0.9994$.
Using the three-dimensional method to analyse lensing by large-scale structure, we find that large magnifications are not produced, as shown in Fig.~\ref{RBMvsMudMag}. 
The slope of the differential magnification probability distribution is found to be much steeper than for RBM for the regime $1.1<\mu<1.5$. The difference in the two methods can also quantified by the standard deviation, found to be $\sigma_{\mu}=0.205$ from the multiple-lens plane approach and $\sigma_{\mu}=0.021$ from the three-dimensional approach, and order of magnitude lower.
To understand the reason of this difference, we also plot the probability distribution for shear, shown in Fig.~\ref{RBMvsMudShear}. While the overall shape of the distribution function is similar as derived by the two methods, the multiple-lens plane method measures more high shear values and less low shear values. 
The projection of matter onto multiple lens-planes may have artificially increased the shearing effect of cosmological lenses. However, the effect of shot-noise in the standard RBM method is identified as another possible reason for the discrepancy. When the three-dimensional approach is used, the gravitational potential is calculated by means of Fourier-methods. In contrast, the multiple-lens plane approach determines the potential field at the location of each ray, but adding the potential due to each simulation particle in the plane individually. Each particle in the simulation represents some underlying smooth mass distribution, but by treating them as point masses, the method is susceptible to the effects of shot-noise.
Recently, \citet{T11} discussed the effects of shot-noise on the variance of the convergence found by ray-tracing through N-body simulations. Although they did not use the same ray-bundle approach that we do here, they apply a multiple lens-plane method with an FFT for calculating the gravitational potential in each plane; they compare their results for a range of Fourier-grid resolutions (see Figure 2 of their paper). When the grid resolution is large enough to smear out structure, the variance of the convergence falls below theoretical predictions; although our choice of statistic is different, our results agree with this interpretation (see Fig.~\ref{nfwtests2} in Sec.~\ref{fakenfw}). Interestingly, a very small grid-resolution ($<5$ h$^{-1}$kpc) leads to a variance that is larger than the theoretical prediction, a result which they attribute to shot-noise. We agree with this interpretation, noting that Fourier methods applied to grid-resolutions that are much smaller than the mean-interparticle distance would resemble a direct summation as employed by the multiple-lens plane approach in the present work.

\begin{figure}
	 \begin{center}  
      	\includegraphics[width=0.95\linewidth]{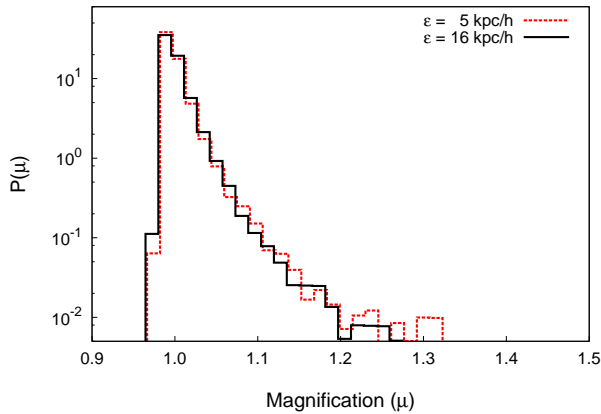}
      	\end{center}
      	\caption{The magnification probability distribution for a source at redshift $z=0.5$ as predicted by the three-dimensional method. Two different force-softening lengths applied in the N-body simulations are compared: $\epsilon = 5$ h$^{-1}$kpc (red dashed line), and $\epsilon = 16$ h$^{-1}$kpc (black solid line)}
	\label{MudMagSoftening}
	\vspace{5mm}
\end{figure}

The $\mu$PDF predicted by the three-dimensional method is subject to certain choices of numerical parameters, some of which have been discussed above. Thus, we test the parameters that are most likely to have an impact on the cosmological lensing results presented above.
For example, a parameter that was not relevant to the tests presented in Sec.~\ref{models} is the force-softening length chosen for the cosmological N-body simulation. We re-run the simulation described earlier but with a force-softening length reduced to $\epsilon = 5$ h$^{-1}$kpc. We sample $3\times10^{4}$ lines of sight through the mass distribution determined with this simulation, and in Fig.~\ref{MudMagSoftening}, we compare the $\mu$PDF to the result for the three-dimensional method shown in Fig.~\ref{RBMvsMudMag}. For each, the Fourier-grid resolution is 49 h$^{-1}$kpc. The similarities in the PDFs reassures us that any smoothing of structure below 16 h$^{-1}$kpc is not responsible for any error in the PDFs measured by the three-dimensional method. 

\begin{figure}
	 \begin{center}  
      	\includegraphics[width=0.95\linewidth]{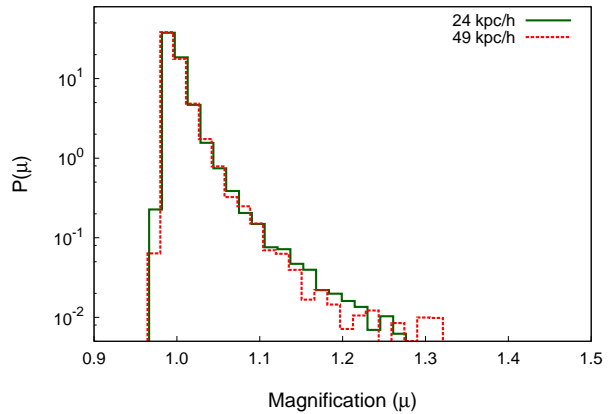}
      	\end{center}
      	\caption{The magnification probability distribution for a source at redshift $z=0.5$ as predicted by the three-dimensional method. Two different Fourier-grid resolutions are compared: 24 h$^{-1}$kpc (green solid line), and 49 h$^{-1}$kpc (red dashed line)}
	\label{MudMagFFTres}
	\vspace{5mm}
\end{figure}

In addition, as the Fourier-grid scale is reduced, the numerical solution is expected to be more accurate, with the caveat that it remains large enough to prevent the effects of shot-noise. In Fig.~\ref{MudMagFFTres} we compare two different Fourier-grid resolutions: 24 h$^{-1}$kpc for which $9\times10^{4}$ lines of sight have been sampled, and 49 h$^{-1}$kpc for which $3\times 10^{4}$ lines of sight have been sampled. Both tests are run on simulations with a force-softening length reduced to $\epsilon = 5$ h$^{-1}$kpc, which are only able to reliably describe structure on scales larger than 20 h$^{-1}$kpc. The differences are negligible, which demonstrate that most of the weak lensing results from structure above 49 h$^{-1}$kpc scales. Structures on scales larger than 24 h$^{-1}$kpc that are able to produce strong or even `moderate' lensing are rare. 

\begin{figure}
	 \begin{center}  
      	\includegraphics[width=0.95\linewidth]{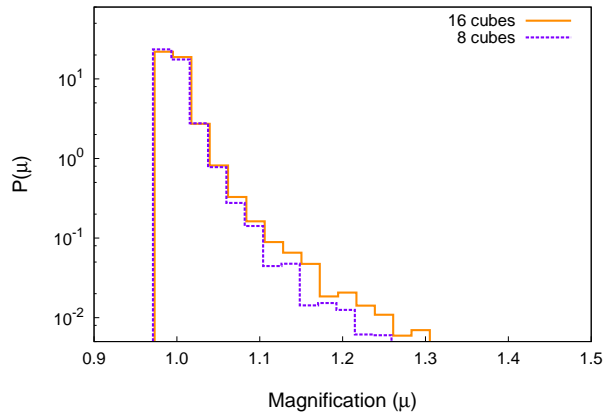}
      	\end{center}
      	\caption{The magnification probability distribution for a source at redshift $z=0.5$ as predicted by the three-dimensional method. The same Fourier-grid resolution (24 h$^{-1}$kpc) is created, but excising either 8 cubes (purple dashed line) or 16 cubes (yellow solid line) from each simulation box}
	\label{MudMagBoxes}
	\vspace{5mm}
\end{figure}

We recognize that there are multiple ways to create the same Fourier-grid resolution. If the number of cubes taken from the simulation boxes are doubled, but the number of Fourier grid-points halved, the resolution does not change. However, only a quarter of the patch of sky would be sampled, and the computational run-time would double. On the upside, the memory usage reduces to an {\it eighth}, which is a significant advantage for such a computational demanding approach. In Fig.~\ref{MudMagBoxes} we show the result of doubling the number of cubes in such a manner; the Fourier-grid resolution is 24 h$^{-1}$kpc. A total of $3\times10^{4}$ lines of sight have been sampled for the method using 8 cubes, while $6\times10^{4}$ lines of sight have been sampled for the method using 16 cubes. Note that this was not tested on the compact lens models in Sec.~\ref{models} since cubes are not excised from simulations as they are when analysing cosmological lensing. We are satisfied that the choice of 8 cubes for our previous results (see Fig.~\ref{RBMvsMudMag} and Fig.~\ref{RBMvsMudShear}) has no bearing on the results.
%

%---------------------------------------------------------------------------------------------------------------------------------

\section{Summary and Discussion}\label{summary}
Modelling the magnification probability distribution of background sources due to gravitational lensing relies on the application of ray-tracing methods. This work has taken the first steps towards making a direct comparison between the predictions of the multiple lens-plane ray-bundle method (RBM) and one that does not invoke the thin-lens approximation; instead, the null geodesic equations are integrated. The efficiency and accuracy of this computationally challenging approach can be improved by careful choices of numerical parameters; therefore, the results are analysed for the behaviour of the ray-tracing code in the vicinity of Schwarzschild and NFW lenses.  A range of tests were able to pin down the numerical parameters that play a critical role in the predicted statistics. The behaviour of the ray-bundles in the vicinity of a Schwarzschild lens demonstrated that the method can reproduce large magnifications given sufficient spatial resolution. The limitations are dominated by the spatial resolution of the Fourier-grid and the mass resolution of the discretized lens. 

Comparisons to a multiple lens-plane algorithm are drawn in the context of cosmological mass distribution for a source redshift of $z_{s}=0.5$. The weak lensing statistics predicted by ray-tracing through simulated cosmological mass distributions found significant differences compared to the results from the original RBM approach. 
Either the use of multiple lens-planes or shot-noise are responsible for the observed differences. 
To clarify the dominant factor, a multiple lens-plane ray-bundle method that applies a two-dimensional Fourier-method at each lens-plane is proposed as the next step towards the direct comparison. 

The method developed here presents a computational challenge as the three-dimensional FFTs prove memory-expensive, but is justified by the need to quantify the multiple lens-plane approximation. We are, as yet, unable to sample the large number of lines of sight required to model the intermediate magnification ($2<\mu<20$) region where \citet{R92} had identified the feature present only in two dimensional lens models.

We now discuss future applications of the this ray-tracing method that are outside the scope of the present work. For example, proposed large surveys have the measurement of cosmic shear, with particular focus on the determination of the nature of dark energy, as one of their main science drivers, using a combination of a large area of sky coverage and high-precision photometric redshifts. These include the ground-based Panoramic Survey Telescope and Rapid Response System (Pan-STARRS)\footnote{http://pan-starrs.ifa.hawaii.edu}, the VST-KIlo-Degree Survey (KIDS)\footnote{http://www.astro-wise.org/projects/KIDS/}, the Dark Energy Survey (DES)\footnote{http://www.darkenergysurvey.org/}, the Large Synoptic Survey Telescope (LSST)\footnote{http://www.lsst.org} as well as the space-based Euclid \citep{L09}\footnote{http://sci.esa.int/euclid} and the Joint Dark Energy Mission (JDEM)\footnote{http://jdem.lbl.gov/ and http://jdem.gsfc.nasa.gov/}, which has recently been rebranded as the Wide-Field InfraRed Survey Telescope (WFIRST). 
The method developed in the present work has the potential to quantify the effects of multiple lens-plane techniques on the theoretical predictions for cosmic shear measurements.
Another alternative to ray-tracing is to retain the three dimensional mass distribution, but assume that the deflections are small enough to satisfy the Born approximation. Here, the matter that is integrated along the (un-deflected) line of sight is solely responsible for the convergence.
Over larger and larger distances, the Born approximation becomes less and less accurate; several studies have found the requirement for corrections in the construction of the shear power spectrum and higher order bispectrum \citep[e.g.][]{VW01,SC06}.
We envisage that the three-dimensional method could be used to make a one-to-one comparison with the Born approximation, just as we have done here with the multiple lens-plane method.

Finally, with a larger number of ray-bundles and greater sky-sampling, we could turn our attention to higher-order statistics. Flexion is the third-order effect in gravitational lensing and is effectively the gradient of the shear component \citep[see][]{GB05}. It has recently been demonstrated that flexion can be modelled using a ray-bundle method \citep{FL11}; this presents an exciting future application of the method, but surely it is even more necessary to justify the use of the multiple lens-plane formalism.

%---------------------------------------------------------------------------------------------------------------------------------
\section*{Acknowledgments}
M.K. acknowledges support from the University of Sydney Faculty of Science Postgraduate Award, as well as a fellowship from the European Commission's Framework Programme 7, through the Marie Curie Initial Training Network CosmoComp (PITN-GA-2009-238356) and would like to thank the Centre for Astrophysics and Supercomputing at the Swinburne University of Technology for their hospitality and for the use of the Green Machine. PL is supported by the Alexander von Humboldt Foundation. This work is undertaken as part of the Commonwealth Cosmology Initiative (www.thecci.org), and funded by the Australian Research Council Discovery Project DP0665574.

%---------------------------------------------------------------------------------------------------------------------------------
\bibliography{MKpaper02}

%---------------------------------------------------------------------------------------------------------------------------------
\appendix

\section{Christoffel Symbols}\label{AppChristoffel}
The expanding weak-field metric presented in Eqn.~\ref{expandingmetric} has the following non-zero Christoffel symbols:
\begin{eqnarray}\label{christoffel}
    \Gamma_{tt}^{x}  & = & \frac{1}{\mathrm{a}^{3}}\phi_{,x}  \notag \\
    \Gamma_{tt}^{y}  & =  &\frac{1}{\mathrm{a}^{3}}\phi_{,y} \notag \\
    \Gamma_{tt}^{z}  & = & \frac{1}{\mathrm{a}^{3}}\phi_{,z}  \notag \\
    \Gamma_{tt}^{t}  & = & -\frac{\dot{\mathrm{a}}}{\mathrm{a}^{2}}\phi  \notag \\
    \Gamma_{tx}^{x}  =  \Gamma_{ty}^{y}  = \Gamma_{tz}^{z}   & = &\frac{\dot{\mathrm{a}}}{\mathrm{a}^{2}}(\mathrm{a}+\phi) \notag \\
    \Gamma_{tx}^{t}  = \Gamma_{yy}^{x}  = \Gamma_{zz}^{x} & = & \frac{1}{\mathrm{a}}\phi_{,x} \notag \\
    \Gamma_{xx}^{x} =  \Gamma_{xy}^{y}  = \Gamma_{xz}^{z} & = &-\frac{1}{\mathrm{a}}\phi_{,x} \notag \\
    \Gamma_{ty}^{t}   = \Gamma_{xx}^{y}  = \Gamma_{zz}^{y}  & = & \frac{1}{\mathrm{a}}\phi_{,y} \notag \\
    \Gamma_{yy}^{y} =  \Gamma_{yx}^{x}  = \Gamma_{yz}^{z} & = &-\frac{1}{\mathrm{a}}\phi_{,y} \notag \\
    \Gamma_{tz}^{t}   =  \Gamma_{xx}^{z}  = \Gamma_{yy}^{z} & = & \frac{1}{\mathrm{a}}\phi_{,z} \notag \\
    \Gamma_{zz}^{z} =  \Gamma_{zx}^{x}  = \Gamma_{zy}^{y} & = &-\frac{1}{\mathrm{a}}\phi_{,z} \notag \\
    \Gamma_{xx}^{t}  = \Gamma_{yy}^{t} = \Gamma_{zz}^{t}     & = & \dot{\mathrm{a}}(\mathrm{a}+\phi)
\end{eqnarray}
Note that $x$ denotes $x^{1}$, $y$ denotes $x^{2}$, and $z$ denotes $x^{3}$. 
The Christoffel symbols for this metric are dependent, not only on the gradients of the gravitational potential $\phi$, but $\phi$ {\it itself}. However, the weak field condition requires that the magnitude of the perturbations satisfy $\phi \ll\mathrm{a}$. The Geodesic Equations - the second order differential equations for the four coordinates - are then constructed.

\section{The Interpolation Scheme}\label{AppInterp}
Here we only describe the scheme for the evaluation of $\phi(x^{1}_{\star},x^{2}_{\star},x^{3}_{\star})$ given the gridded values $\phi(x^{1},x^{2},x^{3})$, but the same method is applied to evaluate the derivatives of the potential. The scheme is not strictly tri-cubic interpolation, but a three-dimensional version of the bicubic spline. Instead, it initially performs a one-dimensional spline to compute the derivative with respect to the line of sight; i.e. it fits a spline to find $d\phi(x^{1},x^{2},x^{3})/dx^{3}$ across the entire grid. This is only done once for each lensing mass distribution.
Each time the local values need to be found, the scheme identifies the gridded values that include $n_{int}$ points on either side of the current location for each dimension, i.e. a $(2n_{int})^{3}$ mesh grid. Then, the scheme performs the following steps:
\begin{enumerate}

\item The gridded derivative is interpolated across the chosen dimension, in this case the $x^{3}$-dimension, to find $\phi(x^{1},x^{2},x^{3}_{\star})$, i.e. the value with the current $x^{3}$-coordinate of the photon specified, but the other coordinates corresponding to grid points. $(2n_{int})^{2}$ interpolations are required.

\item Another set of $2n_{int}$ one-dimensional splines are performed, this time across the $x^{1}$-axis, to find $d\phi(x^{1},x^{2},x^{3}_{\star})/dx^{1}$.

\item $2n_{int}$ interpolations along the $x^{1}$-axis determine $\phi(x^{1}_{\star},x^{2},x^{3}_{\star})$.

\item A single one-dimensional spline across the $x^{2}$-axis is used to find $d\phi(x^{1}_{\star},x^{2},x^{3}_{\star})/dx^{2}$

\item Finally, a single interpolation along the $x^{2}$-axis allows one to evaluate $\phi(x^{1}_{\star},x^{2}_{\star},x^{3}_{\star})$

\end{enumerate}

\section{Schwarzschild Lens}\label{AppSchw}
A single point of infinite density is one of the simplest lens models, and is often used to model a spherically symmetric lens for idealised analytical studies. 
The relevant (angular) scale length for a so-called Schwarzschild lens of mass $M$ is the Einstein radius:
\begin{equation}\label{Eangle}
 \theta_{\mathrm{Ein}} =  \sqrt{\frac{D_{ds}}{D_{d}D_{s}}} \frac{\sqrt{4GM}}{c} .
\end{equation}
At the lens plane, this corresponds to a physical scale length given by:
\begin{equation}\label{Eradius}
 r_{\mathrm{Ein}}  =  \theta_{\mathrm{Ein}} D_{d}.
\end{equation}
The gravitational lensing quantities of interest are found by combining the derivatives of this potential, as described in Sec.~\ref{gravlensing}. They are defined in terms of a dimensionless radial distance, $x=r/r_{\mathrm{Ein}}$. All lines of sight probe regions outside the lens, so the convergence is zero.
The solution for the shear for images lensed by a Schwarzschild lens, $\gamma_{\mathrm{Schw}}$, is:
\begin{equation}\label{schwgamma}
\gamma_{\mathrm{Schw}}(x) = x.
\end{equation}
The analytic solution for the magnification for an image at any location, $\mu_{\mathrm{Schw}}$, is:
\begin{equation}\label{schwmu}
\mu_{\mathrm{Schw}}(x) = \left(1-\frac{1}{x^{4}}\right)^{-1}.
\end{equation}

\section{NFW lens}\label{AppNFW}
The NFW profile, given in Eqn.~\ref{nfwprof}, can be considered a thin lens for cases where the observer, lens and source are separated by large angular diameter distances. The following analytic solutions, \ref{nfwkappa}-\ref{nfwgmore}, are therefore derived from projected surface densities. A dimensionless radial distance, $x=r/r_{s}$, has been adopted. Here, the quantities $\rho_{c}$, $r_{s}$, $\delta_{c}$ and  $\Sigma_{crit}$ are those defined in Eqns.~\ref{critdens}, \ref{nfwprof}, \ref{overdens} and \ref{sigmacrit}. 
Following \citet{WB00} and \citet{C10}, the analytic solution for the radial dependence of the convergence is given by
\begin{equation}\label{nfwkappa}
\kappa_{\mathrm{NFW}}(x) = \frac{r_{s}\delta_{c}\rho_{c}}{\Sigma_{crit}} K(x) ,\end{equation}
where $K(x)$ is given by
\begin{equation}\label{nfwK}
K(x) = \left\{
\begin{array}{rll}
&\frac{2}{(x^{2}-1)}\left[1-\frac{2}{\sqrt{1-x^{2}}}\mathrm{tanh}^{-1}\sqrt{\frac{1-x}{1+x}}\right]    & \text{if } x<1 \\ \\ 
&\frac{2}{3}                         & \text{if } x=1 \\ \\ 
&\frac{2}{(x^{2}-1)}\left[1-\frac{2}{\sqrt{x^{2}-1}}\tan^{-1}\sqrt{\frac{x-1}{1+x}}\right]    & \text{if } x>1
\end{array} \right..
\end{equation}
By integrating Eqn.~\ref{nfwkappa} over the area within $r$, one finds the mass within a {\it cylinder} of radius $r$:
 \begin{equation}\label{nfwcylinder}
M_{\mathrm{NFW,cyl}}(x) = 4\pi r_{s}^{3}\delta_{c}\rho_{c} C(x) ,
\end{equation}
where $C(x)$ is given by
\begin{equation}\label{nfwC}
C(x) = \ln\frac{x}{2} + \left\{
\begin{array}{rll}
&\frac{1}{\sqrt{1-x^{2}}}\mathrm{cosh}^{-1}\frac{1}{x}    & \text{if } x<1 \\ \\ 
&1                         & \text{if } x=1 \\ \\ 
&\frac{1}{\sqrt{x^{2}-1}}\cos^{-1}\frac{1}{x}    & \text{if } x>1
\end{array} \right..
\end{equation}
The radial dependence of shear is thus given by
\begin{equation}\label{nfwgamma}
\gamma_{\mathrm{NFW}}(x) = \frac{r_{s}\delta_{c}\rho_{c}}{\Sigma_{crit}} G(x) ,
\end{equation}
where $G(x)$ is given by
\begin{equation}\label{nfwG}
G(x) = \left\{
\begin{array}{rll}
& g_{<}(x)   	& \text{if } x<1 \\ \\ 
&\left[ \frac{10}{3}+4\ln\left(\frac{1}{2}\right)\right]                        & \text{if } x=1 \\ \\ 
& g_{>}(x)	& \text{if } x>1
\end{array} \right..
\end{equation}
The functions $g_{<}(x)$ and $g_{>}(x)$ are given by
\begin{eqnarray}\label{nfwgless}
g_{<}(x)	&= \frac{8\mathrm{~tanh}^{-1}\sqrt{(1-x)/(1+x)}}{x^{2}\sqrt{1-x^{2}}} + \frac{4}{x^{2}}\ln\left(\frac{x}{2}\right) \notag \\
		&- \frac{2}{(x^{2}-1)} + \frac{4\mathrm{~tanh}^{-1}\sqrt{(1-x)/(1+x)}}{(x^{2}-1)\sqrt{1-x^{2}}}
\end{eqnarray}
and
\begin{eqnarray}\label{nfwgmore}
g_{>}(x) 	& =  \frac{8\tan^{-1}\sqrt{(x-1)/(1+x)}}{x^{2}\sqrt{x^{2}-1}} + \frac{4}{x^{2}}\ln\left(\frac{x}{2}\right) \notag \\
		& - \frac{2}{(x^{2}-1)} + \frac{4\tan^{-1}\sqrt{(x-1)/(1+x)}}{(x^{2}-1)^{3/2}} .
\end{eqnarray}
 Combining these with Eqns.~\ref{nfwkappa} and \ref{nfwgamma} with Eqns.~\ref{poisson} and \ref{magnification} allows one to determine the analytic value of the magnification, $\mu$, of an image at a given projected distance from the centre of the NFW lens.

\section{Full Beam vs Empty Beam}\label{AppBeam}
Magnification is the relative increase in flux as a result of gravitational lensing. There is a subtlety here. One may compare the flux received from the source to the flux received if the universe was entirely homogeneous, in which case the magnification is referred to as the full-beam magnification, $\mu_{fb}$. This is the magnification referred to in Sec.~\ref{mpdf}. The so-called empty-beam magnification, $\mu_{eb}$,  is defined relative to the case where all the matter in the universe is locked up in compact objects, and all lines of sight are empty. As this is the case of minimal magnification (no convergence and assumed negligible shear), then $\mu_{eb}$ is always greater than unity. To derive the relationship between the two magnifications, we must compare the solid angles subtended by the source: $\Omega_{fb}$ in the full-beam scenario, $\Omega_{eb}$ in the empty-beam scenario and $\Omega_{img}$ the solid angle of the image observed. Given the conservation of surface brightness, the magnifications are defined by:
\begin{equation}\label{fullbeammag}
\mu_{fb} = \frac{\Omega_{img}}{\Omega_{fb}}
\qquad \mathrm{and} \qquad
\mu_{eb} = \frac{\Omega_{img}}{\Omega_{eb}}.
\end{equation}
The solid angles are related to the physical area of the source via the angular diameter distances: 
\begin{equation}\label{solidanglesfull}
 \Omega_{fb}(z) = \frac{A_{src}}{(1+z)^{2}D_{fb}(z)}
 \end{equation}
is the appropriate relationship for the full-beam case, and
\begin{equation}\label{solidanglesempty}
 \Omega_{eb}(z) = \frac{A_{src}}{(1+z)^{2}D_{eb}(z)}.
\end{equation}
is appropriate for the empty-beam case.
Combining Equations \ref{fullbeammag}, \ref{solidanglesfull} and \ref{solidanglesempty}, one is easily able to convert between the two:
\begin{equation}\label{comparemag}
\frac{\mu_{fb}}{\mu_{eb}} = \frac{D_{fb}^{2}(z)}{D_{eb}^{2}(z)}.
\end{equation}
The angular diameter distance in the full-beam case, $D_{fb}(z)$, is equivalent to the  Dyer-Roeder distance $D(\tilde{\alpha}=1; z)$ which is also the FLRW solution:
\begin{equation}\label{angularDD12}
    D_{fb}(z) \equiv  \frac{c}{H_{0}(1+z)} \int^{z}_{0}\frac{dz'}{\left[\OmMo(1+z')^{3}+ \OmLo \right]^{1/2}}.
\end{equation}
The empty-beam angular diameter distance, $D_{eb}(z)$, is found by solving the Dyer-Roeder equation for $\tilde{\alpha}=0$ : 
\begin{equation}
D_{eb}(z) \equiv D(\tilde{\alpha}=0; z) = c \int^{z}_{0} \frac{1}{(1+z')^{2}H(z')} dz'
\label{DRempty}
\end{equation}

\bsp

\label{lastpage}
\end{document}